\documentclass{emulateapj}

\usepackage{graphicx}
\usepackage{natbib}
\usepackage{longtable,lscape}
\slugcomment{Astrophysical Journal, in press}
\shorttitle{Spectroscopy of X-ray sources in the ECDF-S}
\shortauthors{Treister et al.}

\begin{document}
\title{Optical Spectroscopy of X-ray sources in the Extended Chandra Deep Field South\footnote{This paper includes data gathered with the 6.5 meter Magellan Telescopes located at Las Campanas Observatory, Chile.}~~\footnote{Partly based on observations collected at the European
Southern Observatory, Chile, under programs 078.A-0485 and 072.A-0139.}}

\author{Ezequiel Treister\altaffilmark{1,2,3}, Shanil Virani\altaffilmark{4}, Eric Gawiser\altaffilmark{5}, C. Megan Urry\altaffilmark{4,6}, Paulina Lira\altaffilmark{7},  Harold Francke\altaffilmark{7}, Guillermo A. Blanc\altaffilmark{8}, Carolin N. Cardamone\altaffilmark{4}, Maaike Damen\altaffilmark{9}, Edward N. Taylor\altaffilmark{9} and Kevin Schawinski\altaffilmark{4}}

\altaffiltext{1}{Institute for Astronomy, 2680 Woodlawn Drive, University of Hawaii, Honolulu, HI 96822; treister@ifa.hawaii.edu}
\altaffiltext{2}{European Southern Observatory, Casilla 19001, Santiago 19, Chile.}
\altaffiltext{3}{Chandra Fellow}
\altaffiltext{4}{Department of Astronomy, Yale University, PO Box 208101, New Haven, CT 06520.}
\altaffiltext{5}{Department of Physics and Astronomy, Rutgers, 136 Frelinghuysen Road, Piscataway, NJ 08854-8019.}
\altaffiltext{6}{Department of Physics, Yale University, P.O. Box 208121, New Haven, CT 06520.}
\altaffiltext{7}{Departamento de Astronom\'{\i}a, Universidad de Chile, Casilla 36-D, Santiago, Chile.}
\altaffiltext{8}{Department of Astronomy, University of Texas at Austin,  1 University Station, C1400 Austin, Texas 78712}
\altaffiltext{9}{Sterrewacht Leiden, Leiden University, NL-2300 RA Leiden, Netherlands.}

\begin{abstract} 
We present the first results of our optical spectroscopy program aimed to provide
redshifts and identifications for the X-ray sources in the Extended Chandra Deep Field
South. A total of 339 sources were targeted using the IMACS spectrograph at the Magellan
telescopes and the VIMOS spectrograph at the VLT. We measured redshifts for 186 X-ray
sources, including archival data and a literature search. We find that the AGN host
galaxies have on average redder rest-frame optical colors than non-active galaxies, and
that they live mostly in the ``green valley''. The dependence of the fraction of AGN that
are obscured on both luminosity and redshift is confirmed at high significance and the
observed AGN space density is compared with the expectations from existing luminosity
functions. These AGN show a significant difference in the mid-IR to X-ray flux ratio for
obscured and unobscured AGN, which can be explained by the effects of dust self-absorption
on the former. This difference is larger for lower luminosity sources, which is consistent
with the dust opening angle depending on AGN luminosity.
\end{abstract}

\keywords{galaxies: active -- quasars: general -- X-rays: galaxies}

\section{Introduction}

X-ray surveys have been critical in unveiling a large population of obscured supermassive
black holes actively accreting matter (e.g., \citealp{giacconi01} and references therein),
contrary to optical surveys, like the Sloan Digital Sky Survey (e.g.,
\citealp{schneider02}), which are dominated by unobscured Active Galactic Nuclei (AGN). 
The hard spectral shape of the X-ray Background (XRB; e.g. \citealp{gruber92}) reveals
that obscured AGN should significantly outnumber the unobscured sources, as was first
proposed by \citet{setti89}. Given the difficulties in detecting and identifying the low
luminosity and/or high redshift AGN, in particular those heavily obscured, it has been
very challenging to obtain a complete census of the AGN population. This is particularly
important in order to study the growth of supermassive black holes and its relation to the
formation and evolution of its host galaxy.

Deep X-ray surveys like the Chandra Deep Fields North (CDF-N;
\citealp{brandt01}) and South (CDF-S; \citealp{giacconi01,rosati02})
have been critical in obtaining a significant sample of obscured AGN
up to significant distances. The first results from the optical
follow-up of the X-ray sources in the Chandra Deep Fields reported a
peak of the AGN activity at $z$$\sim$0.7 \citep{szokoly04}, in
contradiction with the predictions of the early XRB population
synthesis (e.g., \citealp{comastri95,gilli01}), which expected a
redshift peak at $z$$\sim$1.5. This discrepancy was resolved in the
XRB population synthesis models of \citet{treister05b} and later by
\citet{gilli07}, both of whom incorporated the latest Chandra and
XMM-Newton observational results in their calculations. Similarly, the
observed fraction of obscured to unobscured AGN in the Chandra Deep
Fields was $\sim$1-2:1, lower than the value of $\sim$3-4:1 expected
from XRB population synthesis models and observed in samples of nearby
sources \citep{risaliti99}. This was explained by \citet{treister04}
as due to the difficulty in obtaining spectroscopic identifications
for obscured AGN at significant redshifts, which are much fainter in
the optical than the unobscured sources.

One significant problem of these deep multiwavelength surveys is
caused by the effects of cosmic variance. For example, \citet{gilli05}
found a factor of two difference between the correlation length for
AGN derived in the CDF-N and CDF-S fields, each covering
$\sim$0.1~deg$^2$, and concluded that this difference could be
explained by the effects of cosmic variance. In addition, significant
overdensities and large scale structures have been found in both
fields (e.g., \citealp{gilli03}), which potentially affect the
conclusions derived from each field individually. In order to address
this problem, recently several wide-field deep surveys have been
carried out. Examples of these surveys are the All-wavelength Extended
Groth strip International Survey (AEGIS; \citealp{davis07}), the
Cosmic Evolution Survey (COSMOS; \citealp{scoville07}) and the
Multiwavelength Survey by Yale-Chile (MUSYC;
\citealp{gawiser06a,treister07}). The Extended Chandra Deep Field
South (ECDF-S) is one of the four 30$'$$\times$30$'$ fields studied by
MUSYC. This field is $\sim$3$\times$ larger than the existing CDF-N or
CDF-S. Hence, while the effects of cosmic variance are reduced
compared to the results of each of these deep fields, the uncertainty
due to large scale structure can still be relevant for our results. To
minimize these effects, when possible, we combine in this work our
data with available measurements in other fields.

Deep optical imaging has been obtained in the ECDF-S, mostly using the
Wide Field Imager (WFI) on the 2.2-metre telescope at La Silla. Images
in a narrow band filter centered at 500 nanometers were also obtained
in this field in order to look for Lyman Alpha Emitters at $z$$\sim$3
\citep{gawiser06b}.  The ECDF-S imaging was made public by the ESO
Deep Public Survey \citep{mignano07}, COMBO-17 \citep{wolf04}, and
Garching-Bonn Deep Survey (GaBODS; \citealp{hildebrandt06})
teams. Deep near-IR coverage was obtained using the CTIO 4-metre
telescope with the Infrared Sideport Imager (ISPI), reaching a
magnitude limit of $\sim$22 (AB) in the $JHK_s$ bands\footnote{Optical
and near-IR images and catalogs are public and can be obtained at
http://www.astro.yale.edu/MUSYC.}  (E. Taylor et al. in prep.). In
addition, the ECDF-S is the target of two Spitzer legacy surveys, the
Spitzer IRAC / MUSYC Public Legacy in ECDF-S (SIMPLE) and the
Far-Infrared Deep Extragalactic Legacy Survey (FIDEL), which obtained
deep mid-IR data from 3 to 70 microns. Most relevant for our work is
the Chandra coverage of the ECDF-S, consisting of 4 ACIS-I pointings
centered on the original CDF-S to a depth of $\simeq$230~ksec and
covering an area of $\simeq$0.3~deg$^2$. A detailed description of
these data was presented by \citet{lehmer05} and \citet{virani06}. In
this work we use the \citet{virani06} catalog for the optical
follow-up. A total of 651 sources are included in this catalog to a
flux limit of 1.7$\times$10$^{-16}$~erg~cm$^{-2}$s$^{-1}$ in the 0.5-2
keV band. These sources constitute the main sample of this work.

In this paper we present the first results of an extensive
spectroscopic effort aimed to identify a significant number of the
X-ray sources in the ECDF-S field. Optical spectroscopy allow us to
measure redshifts, and hence luminosities, and also to provide an
indication of the nature of the source, which can be compared with the
observed multiwavelength properties. In addition, we will focus on the
properties of the obscuring material by studying the mid-IR properties
of these sources, and by measuring the fraction of obscured AGN and
its dependence on parameters like luminosity or redshift. In Section 2
we present the spectroscopic data and outline the reduction techniques
used in this work, while in \S 3 the results are presented. In Section
4 we take advantage of the exquisite multiwavelength coverage
available on the ECDF-S to compare the X-ray, optical and mid-IR
properties of these sources. We also study the statistical properties
of these sources in order to understand the evolution and physical
conditions of the AGN population. The conclusions of our work are
presented in \S 5. When required, we assume a $\Lambda$CDM cosmology
with $h_0$=0.7, $\Omega_m$=0.3 and $\Omega_\Lambda$=0.7, in agreement
with the most recent cosmological observations \citep{spergel07}. All
magnitudes in this paper are presented in the AB photometric system
\citep{oke83}.

\section{Optical Spectroscopy}

Sources targeted for spectroscopy were selected from the catalog of Chandra sources of
\citet{virani06}. We use this catalog instead of a similar one presented by
\citet{lehmer05} since the former uses a more detailed rejection of high-background
periods, thus eliminating spurious sources. The \citet{virani06} catalog consists of 651
X-ray detected sources, 587 detected at high significance (the primary catalog) and 64
found using a lower threshold (secondary catalog). The X-ray sources were then matched to
the MUSYC BVR optical catalogue presented by \citet{gawiser06b} using a likelihood
ratio technique, as described by \citet{cardamone08}, using a reliability threshold of 0.6. The
estimated number of false matches using this technique, calculated by randomly changing
the X-ray positions, is $\sim$1\%. We found optical counterparts for 445 X-ray sources in
the primary catalog (76\%) and 28 in the secondary catalog (44\%) down to a limiting
magnitude of BVR=27.1. Using the Spitzer IRAC observations of the ECDF-S described in
\S2.5 and a similar likelihood ratio method we found near-IR counterparts for 554 (94\%)
X-ray sources in the primary catalog and 43 (67\%) in the secondary catalog. Hence, we are
very confident that a very large fraction of the X-ray sources, $>$90\% in the primary and
$\sim$70\% in the secondary catalogs, are real.

While in principle all X-ray sources with optical counterparts were targeted for
spectroscopy, we gave higher priority to the optically brightest sources, in order to
increase the efficiency of our observations. In general, sources with $R<$24 had a higher
priority, except for our VIMOS (VLT Visible Multi-Object Spectrograph) observations, in
which an $R$$<$25 threshold was used. A total of 235 sources with $R$$<$24 and 292 with
$R$$<$25 were targeted. As an experiment, for our IMACS (Inamori Magellan Areal Camera and
Spectrograph) run of November 2005 we gave a higher priority to a group of 70 X-ray
sources with bright optical counterparts ($R<$24) and a hard X-ray spectrum, given by a
hardness ratio (HR) lower than -0.2. This was done in order to see if these sources have
signs of obscuration in the optical spectrum as well, and to try to partially overcome the
bias against obscured sources in identification studies based in optical spectroscopy. All
these possible selection effects are considered on the analysis presented in
\S\ref{discussion}.

\subsection{Magellan IMACS data}

We used the IMACS wide-field spectrograph mounted on the Magellan I Baade telescope at Las
Campanas Observatory. This camera provides a circular field of view with a 27.2$'$ radius
in the $f/2$ configuration. The pixel scale for this setup is 0.2$''$/pixel. For all our
observations we used the 300 lines/mm grism and slits of 1.2$''$ width, except for the
October 2003 and 2006 runs in which we used 1$''$ slits. For our average seeing of 1$''$,
this translates into resolution elements of 8\AA~ (6.5\AA~ for the October 2003 and 2006
runs) except for the November 2005 run in which the average 0.6$''$ seeing means a
resolution element of $\sim$4\AA~ for unresolved sources.  While the wavelength coverage
depends strongly on the position of the source in the mask, we required a minimum
wavelength coverage of 4200-7000\AA~, which maximizes the number of sources on a given
mask while at the same time providing a broad wavelength coverage, which is important
considering that the redshifts are not know a priori and can span a wide range. The log of
our ECDF-S IMACS observations is presented in Table~\ref{log_imacs}.

Typically each mask was observed for 5 hours. The goal was to reach a magnitude of
R$\sim$24, however in many cases, due to weather or technical problems, this was not
possible. Hence, the efficiency, defined as the fraction of sources identified and with a
measured redshift, differs significantly from mask to mask. The total number of sources on
a typical mask is $\sim$80-100, of which 15-50\% are X-ray sources. The large spread in
the fraction of X-ray sources on a given mask reflects the fact that these sources were
not always the main target of the observations, but ``mask-fillers''. The remaining
targets include extragalactic sources like Ly-$\alpha$ emitters, Lyman break galaxies and
galactic white and brown dwarf stars selected by their optical colors from the MUSYC
catalogues. The properties of the general sources targeted with IMACS will be presented in
a following paper (P. Lira et al. in prep.).

\subsection{VLT VIMOS data}

The VIMOS \citep{lefevre03} spectrograph mounted at the Nasmyth focus B of UT3 was also
used to obtain optical spectroscopy for the X-ray sources in the ECDF-S. This instrument
consists of four cameras of 7$'$$\times$8$'$ each, hence providing a total field of view
of 15$'$$\times$16$'$, with a pixel scale of 0.2$''$/pixel. We used the MR grism together
with the OS-blue filter in order to block the contamination from higher orders. In this
setup the wavelength coverage is 4000-6700\AA~, very similar to the coverage of the IMACS
observations. This coverage minimizes the effects of fringing at red wavelengths and takes
advantage of the wavelength region of maximum efficiency of the VIMOS camera. The MR grism
provides a dispersion of 2.5\AA~/pixel, which combined with our 1$''$ wide slits
corresponds to a resolution element of 12.5\AA~. All the VIMOS observations were carried
out in service mode by the Paranal staff. The required conditions were seeing better than
1$''$, clear sky and fractional lunar illumination smaller than 20\% at a minimum distance
of 30 degrees. Due to the lack of an Atmospheric Dispersion Corrector in VIMOS and in
order to minimize the effects of instrument flexures, observations were carried out at a
maximum hour angle of $\pm$2 hours.

Four pointings were used to completely cover the ECDF-S 30$'$$\times$30$'$ field. Each
mask was observed for three hours and 18 minutes, which corresponds to a total execution
time of six hours including overheads. According to the exposure time calculator this
should allow us to significantly detect sources up to R$\sim$24.5. A total of 283 sources
were targeted, excluding alignment stars, 96 of them X-ray sources.

\subsection{Additional Observations}

Since the ECDF-S is the target of many deep multiwavelength observations, it is not
surprising that many spectroscopic follow-up programs have been carried out in this
field. In order to take advantage of existing observations of these X-ray sources, we did
an extensive literature search. In particular, we used the Master Compilation of
GOODS/CDF-S spectroscopy\footnote{Available at
http://www.eso.org/science/goods/spectroscopy/CDFS\_Mastercat/}. We matched this
spectroscopic catalog to our MUSYC optical catalog using a search radius of 0.7$''$. Then,
we used the MUSYC-Chandra matched catalog described before and found spectroscopic
identifications for the X-ray sources in the observations reported by
\citet{vanzella05,vanzella06,vanzella08}, \citet{lefevre04}, \citet{szokoly04} and
\citet{croom01}. In total, existing identifications of 55 X-ray sources were added to our sample.

We also searched for existing unpublished spectroscopic observations in the ESO-VLT
archive\footnote{The ESO archive can be found at http://archive.eso.org}. We focused on
VIMOS, given the large area covered by this camera and that most of the existing FORS2
observations were already incorporated in the master compilation described above. There
are several VIMOS observations of the ECDF-S on the ESO archive, however of particular
relevance for us was the program of Bergeron et al. (program ID 072.A-0139), for which 23
hours were granted in order to obtain spectroscopy of the X-ray sources detected by
XMM. The area covered by XMM is well matched to the ECDF-S Chandra observations, so it is
reasonable that a significant number of our Chandra sources were identified as part of
this program. Three fields covered by the Bergeron program overlap with the ECDF-S. These
fields were observed using both the MR and LR-Blue grisms with the GG475 and OS-Blue
filters respectively. We looked for counterparts of our Chandra sources using a search
radius of 2$''$, which was chosen in order to account for the XMM positional uncertainty
but at the same time avoid mis-identifications. In total, 36 sources in our Chandra
catalog were observed as part of the Bergeron program; 28 of them were already targeted as
part of our IMACS and VIMOS program, but only 11 successfully identified, so potentially
identifications for 25 new sources could be obtained from the archival data.

\subsection{Data Reduction}

The IMACS data were reduced using the Carnegie Observatories System for MultiObject
Spectroscopy (COSMOS) package\footnote{This software can be obtained from
http://www.ociw.edu/Code/cosmos} developed by Gus Oemler. This software was specifically
designed to reduce IMACS data, in particular handling the challenge of having spectra
spread over several CCDs. We followed the standard reduction cookbook, which consists of
the following steps: construction of the spectral map using arc frames, bias-subtraction
and flat fielding of science data and sky subtraction. We used COSMOS to generate 2D
sky-subtracted wavelength-calibrated spectra of each source. We then used IRAF's apall task
in order to extract the 1D spectrum.

To reduce the VIMOS data we used the ESO VIMOS pipeline version 2.1.6\footnote{This
package can be obtained from
http://www.eso.org/projects/dfs/dfs-shared/web/vlt/vlt-instrument-pipelines.html}. In this
case all the spectra for a given quadrant are located on one CCD, so the data reduction is
significantly simpler than for IMACS data. However, the basic reduction steps are very
similar to the IMACS reduction. The same procedure was used to reduce both the archival
observations and our proprietary data. In the case of the VIMOS data from our program, for each
individual observation of $\sim$33 minutes, the corresponding night-time arc was used to
calculate the spectral map, in order to account for the effects of instrument flexures. In
addition, observations of standard stars on the same night were used to calculate the
effective response as a function of wavelength for these observations.

\subsection{Spitzer Data}

In order to study the IR to X-ray ratio for our sample of sources with measured
spectroscopic redshifts in the ECDF-S we took advantage of the deep Spitzer images
available in this field. At the shortest wavelengths, the central CDF-S region
(10$'$$\times$16$'$) was observed by IRAC \citep{fazio04a} as part of the GOODS survey to
flux limits of 0.13, 0.22, 1.45 and 1.61~$\mu$Jy in the 3.6, 4.5, 5.7 and 8~$\mu$m band
respectively. Source matching and basic IR properties were reported by
\citet{treister06a}. In addition, the whole ECDF-S field was covered by IRAC as part of
the SIMPLE survey\footnote{Further information about this survey, catalogs and images can
be found at the website http://www.astro.yale.edu/dokkum/simple/} (Damen et al. in
prep.). The flux limits for the SIMPLE observations are 0.76, 0.4, 5.8 and 3.6 $\mu$Jy in
the 3.6, 4.5 and 5.7 and 8~$\mu$m bands respectively, thus $\sim$3-5$\times$ shallower
than the GOODS observations. The matched catalog of the ECDF-S X-ray sources to the SIMPLE
images used here was presented and described in detail by \citet{cardamone08}.

At longer wavelengths, we use the available Spitzer MIPS \citep{rieke04} observations at
24~$\mu$m. In the central region, GOODS provide deep images to a flux limit of
$\sim$25~$\mu$Jy. A catalog\footnote{More details about this catalog and data-reduction
procedures can be found at the webpage
http://data.spitzer.caltech.edu/popular/goods/Documents/goods\_dataproducts.html} based on
PSF-fitting magnitudes, described by \citet{treister06a} and R. Chary et al. (in prep.),
was matched to the X-ray sources in the GOODS region using the IRAC-derived positions.

\begin{figure}
\begin{center}
\plotone{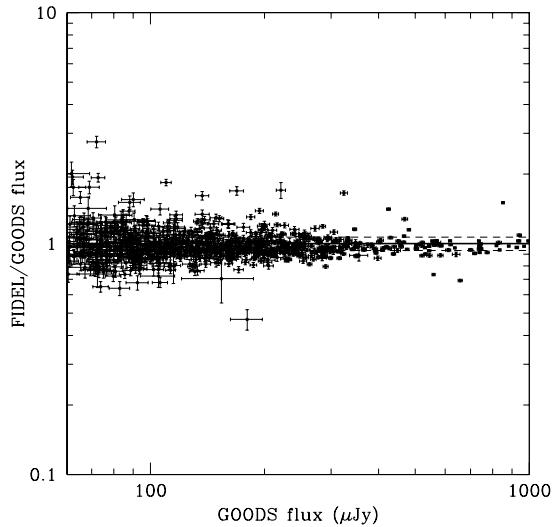}
\end{center}
\caption{Ratio of PSF-fitting 24~$\mu$m fluxes derived from the FIDEL images 
and GOODS catalogs as a function of GOODS flux for the sources significantly detected by
both surveys. The median value for the flux ratio is 0.97, with a standard deviation of
7\%, well within the measurement errors, showing that the two studies produced
consistent results.}
\label{comp_goods}
\end{figure}

Additional data at 24~$\mu$m covering the whole ECDF-S field were taken as part of the
FIDEL survey. Unfortunately, at the moment only the reduced images were made available by
the FIDEL group\footnote{The reduced images and more details about the survey and
reduction techniques can be found at
http://data.spitzer.caltech.edu/popular/fidel/2007\_sep17/fidel\_dr2.html}. In order to
produce our own catalog based on the reduced FIDEL images we used the Mosaicking and
Point-Source Extraction (MOPEX; \citealp{makovoz05}) package, provided by the Spitzer
Science Center. Briefly, we did a first-pass extraction using the default point response
function (PRF) included in MOPEX in order to obtain a preliminary catalog of the brightest
sources. These sources were then used to construct a customized PRF, based directly in the
FIDEL observations. Finally, using this PRF we performed a second extraction and measured
PRF-fitting flux densities for the detected sources. This procedure is described in more
detail by \citet{makovoz05}. In order to verify the accuracy of our derived fluxes in
Figure~\ref{comp_goods} we compare the FIDEL and GOODS sources for the 829 overlapping
sources with fluxes higher than 60~$\mu$Jy in order to ensure a reliable detection in the
FIDEL image. There is a very good agreement between the FIDEL and GOODS fluxes. The median
ratio of FIDEL to GOODS flux is 0.97 with a standard deviation of 7\%. Given the very
small systematic offset between the FIDEL and GOODS fluxes, well within the measurement
errors, no correction to the FIDEL fluxes was applied. The FIDEL catalog was matched to
the ECDF-S using the maximum likelihood procedure described by \citet{cardamone08}. 

\section{Results}
\label{results}

\begin{figure}
\begin{center}
\plotone{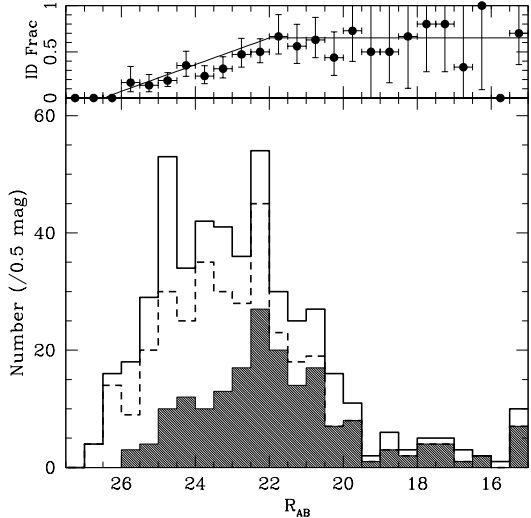}
\end{center}
\caption{{\it Bottom panel}: Distribution of $R$-band magnitude for all X-ray sources with
optical counterpart in the MUSYC images ({\it solid line}), sources
targeted for spectroscopy ({\it dashed}) and identified sources ({\it
hatched histogram}). {\it Top panel}: Fraction of identified to total
X-ray sources as a function of R-band magnitude. The {\it thin solid
line} shows the fit to the data points, used in \S 4. The fraction of
identified sources is $\sim$65\% at $R$$<$22, and decreases towards
fainter optical fluxes.}
\label{r_dist}
\end{figure}

A total of 339 X-ray sources in the \citet{virani06} catalog were targeted for
spectroscopy using the VIMOS and IMACS spectrographs and data from the literature. For 186
of them, we were able to identify the nature of the source and measure the redshift,
giving an average efficiency of 55\%. Of those, 180 were detected in the high-significance
primary X-ray catalog. The basic X-ray, optical and spectroscopic properties of these
sources are presented in Table~\ref{cat}. The R-band magnitude distributions for all the
X-ray sources, the sources targeted for spectroscopy and those with spectroscopic
identification are presented in Figure~\ref{r_dist}. The fraction of sources that are
successfully identified depends strongly on the observing conditions, achieved exposure
time, etc. (Table~\ref{log_imacs}). At the same time, this fraction always decreases
strongly towards fainter magnitudes (Figure~\ref{r_dist}). In our analysis, we model this
dependence using a very simple linear fit, with a constant value of 65\% to $R$=22 and
then decreasing linearly to 0\% at $R$=26.5. The resulting fit is shown in the upper panel
of Figure~\ref{r_dist}.

\begin{figure}
\begin{center}
\plotone{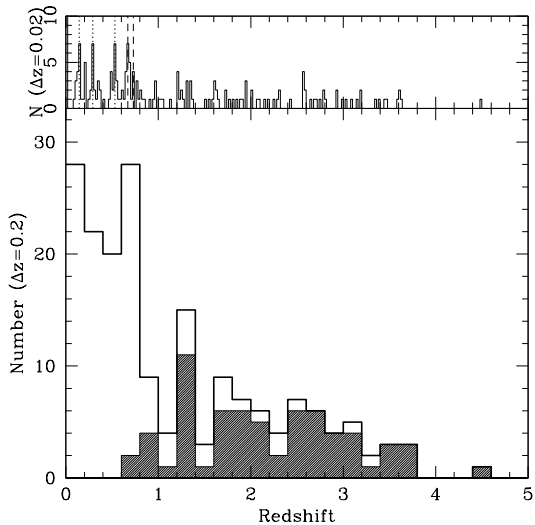}
\end{center}
\caption{{\it Bottom panel:} Redshift distribution for all X-ray sources with spectroscopic
identification ({\it solid line}) and sources classified as unobscured AGN only ({\it
hatched histogram}) in bins of $\Delta z$=0.2. {\it Top panel:} Redshift distribution for
all sources in bins of $\Delta z$=0.02, in order to identify overdensities. The {\it
dashed lines} mark the previously known structures in this field at $z$=0.67 and 0.73,
while the {\it dotted lines} show other suspected overdensities at $z$=0.15, 0.29 and
0.53.}
\label{red_dist}
\end{figure}

The redshift distribution for our spectroscopically-identified sources is presented in
Figure~\ref{red_dist}. The vast majority of unobscured sources (that present broad
emission lines, see discussion below) are at $z$$>$1, with an average redshift of 2.22 and
a median of 2.12. In contrast, obscured AGN and X-ray emitting galaxies are preferentially
located at $z$$<$1. For the total sample, the average redshift is 1.18, with a median of
0.75, in good agreement with the results found in both Chandra Deep Fields
\citep{szokoly04,barger03,gilli05}. The highest redshift source in our sample is a broad
line AGN, XID 213, at $z$=4.48.

\begin{figure}
\begin{center}
\plotone{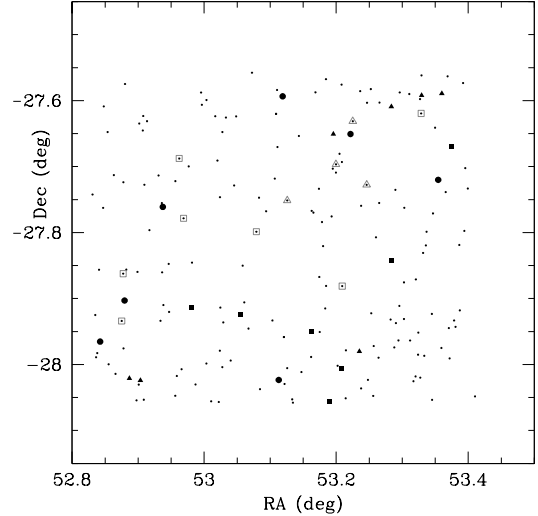}
\end{center}
\caption{{\it Solid small circles}: location of identified sources in the ECDF-S field. Other symbols
mark the sources in the identified overdensities: {\it Solid triangles}: $z$=0.15, {\it
solid squares}: $z$=0.29, {\it large solid circles}: $z$=0.53, {\it open squares}: $z$=0.67,
{\it open triangles}: $z$=0.73. As can be seen, sources in the groups in redshift space
span almost the whole studied area.}
\label{field}
\end{figure}

In the upper panel of Figure~\ref{red_dist} we use a narrower bin width ($\Delta
z$=0.02), in order to search for structures in redshift space. The two main structures,
``walls'', previously detected in the CDF-S proper at $z$=0.67 and $z$=0.73
\citep{gilli03,adami05} are also easily found in our sample. In addition, we found excesses
of sources at $z$=0.15,0.29 and 0.53. In Figure~\ref{field} we present the spatial
distribution of all the identified sources, with the position of the sources in those
redshift structures highlighted. For $z$=0.15,0.29 and 0.53 the 30$'$$\times$30$'$ ECDF-S
field corresponds to a comoving area of 4.6$\times$4.6, 7.7$\times$7.7 and
11.2$\times$11.2 Mpc$^2$ respectively. Given that all the redshift groups span almost the
whole field of view, these structures are much larger than the compact structures reported
by \citet{adami05}. In order to estimate the probability of having seven sources in one
$\Delta z$=0.02 bin we assume that the redshift distribution follows a Poisson
distribution. The average number of sources in a $\Delta z$=0.02 bin is $\simeq$2. Hence,
the probability of having seven sources in one bin randomly is $\simeq$0.3\%. Therefore,
we can conclude that these groups in redshift space should correspond to real physical
associations.

Sources were mainly classified according to their optical spectroscopy
characteristics. Our primary division comes from the presence or absence of broad emission
lines. The threshold for unobscured sources was set at $\Delta$v$>$1000~km~s$^{-1}$,
since \citet{barger05} showed that the vast majority of the non-X-ray sources have
line-widths smaller than this value, while sources with broad lines have also soft X-ray
spectra, consistent with being dominated by an unobscured AGN. We then used the hard (2-8
keV) X-ray luminosity to separate normal galaxies with X-ray emission from AGN-dominated
sources. Specifically, we assumed a conservative and typical threshold of
$L_X$=10$^{42}$erg~s$^{-1}$, higher than the highest level of X-ray activity observed in
star-forming galaxies (e.g., \citealp{lira02}).

\begin{figure}
\begin{center}
\plotone{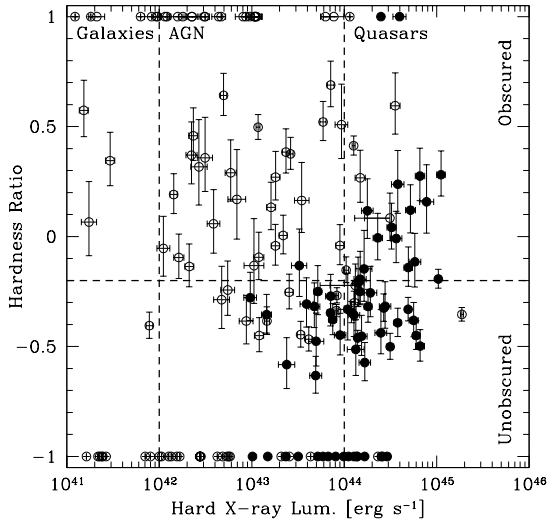}
\end{center}
\caption{Hardness ratio, defined as (H-S)/(H+S) where H and S are the hard (2--8~keV) and 
soft (0.5--2~keV) X-ray count rates, as a function of hard X-ray luminosity for sources
optically classified as obscured ({\it open circles}) and unobscured ({\it filled
circles}) AGN. Vertical dashed lines show the typical separation for normal galaxies
($L_X$$<$10$^{42}$~erg~s$^{-1}$), Seyfert galaxies ($L_X$=10$^{42-44}$~erg~s$^{-1}$) and
quasars ($L_X$$>$10$^{44}$~erg~s$^{-1}$). The dashed horizontal line shows the HR for a
source with $\Gamma$=1.9 and $N_H$=10$^{22}$cm$^{-2}$ at $z$$<$0.5. Most sources
classified as unobscured AGN are clustered in the high luminosity and small HR region,
while obscured AGN have in general lower luminosities and higher HR. }
\label{hr_lum}
\end{figure}

The X-ray spectral shape can be used to obtain an alternative source classification. In
particular, the hardness ratio, defined as the ratio between the difference of the hard
band (typically 2-8 keV for Chandra observations) and soft band (0.5-2 keV) count rates
and their sum is used when the number of detected counts is not enough to perform X-ray
spectral fitting. Figure~\ref{hr_lum} shows the hardness ratio as a function of luminosity
for the 186 sources with spectroscopic identifications presented in this paper. As can be
seen, sources with hard X-ray spectrum (positive hardness ratio) tend to be also optically
classified as obscured sources in the optical, while sources that present broad emission
lines have in general a soft X-ray spectrum (negative hardness ratio). It is also
interesting that most of the unobscured sources, in both X-rays and optical wavebands,
have a high X-ray luminosity ($L_X>$10$^{43}$~erg~s$^{-1}$), while most obscured sources
are fainter in X-rays. These trends will be studied in further detail in the following
section.

While the hardness ratio can be used as a rough indication of the
X-ray spectral shape, for the brighter X-ray sources it is possible to
perform more accurate spectral fitting. In the case of the ECDF-S, a
total of 184 sources have more than 80 counts detected in the
0.5-8~keV Chandra band, making spectral fitting possible. We used a
modified version of the Yaxx\footnote{Available at
http://cxc.harvard.edu/contrib/yaxx/} software to extract the spectrum
of each high-significance source. For sources with more than 200
background-subtracted counts and measured redshifts, 72 in the ECDF-S,
we simultaneously fitted a power-law spectrum and photoelectric
absorption, with three free parameters: slope of the power law
($\Gamma$), normalization and observed neutral hydrogen column density
($N_H$). In all cases we included the Galactic absorption, assuming an
average $N_H$ of 6.8$\times$10$^{19}$~cm$^{-2}$ for the ECDF-S, as
measured by the survey of Galactic HI of \citet{kalberla05}. For
sources with less than 200 but more than 80 counts it was not possible
to perform such detailed fitting, so we fixed the slope of the
power-law to a value of $\Gamma$=1.9, very similar to the average
value for the sources with a fitted slope ($\Gamma$=1.95), which also
corresponds to the typical X-ray spectrum for unobscured AGN
(\citealp{nandra94,nandra97} and others). In both cases, the $N_H$
value was derived in the observed frame, in order to keep the X-ray
spectral fitting independent of the redshift measurements. In order to
calculate the conversion from observed to intrinsic $N_H$ we used the
Xspec version 12.4.0 software \citep{arnaud96} to simulate a X-ray
spectrum with $\Gamma$=1.9 as observed at the Chandra aimpoint with
ACIS-I, added absorption at varying redshifts from 0 to 4 and measured
the fitted $N_H$ value in the observed frame. We obtained an empirical
relation of $N_H$(intr)=(1+$z$)$^{2.65}$$N_H$(obs), consistent with
the conversion assumed in the past for Chandra data (e.g.,
\citealp{bauer04}). Furthermore, we checked that the derived relation
is roughly independent of the assumed slope of the power-law
spectrum. Hence, we are very confident that our conclusions are not
affected by our choice of measuring absorption in the observed frame.

\begin{figure}
\begin{center}
\plotone{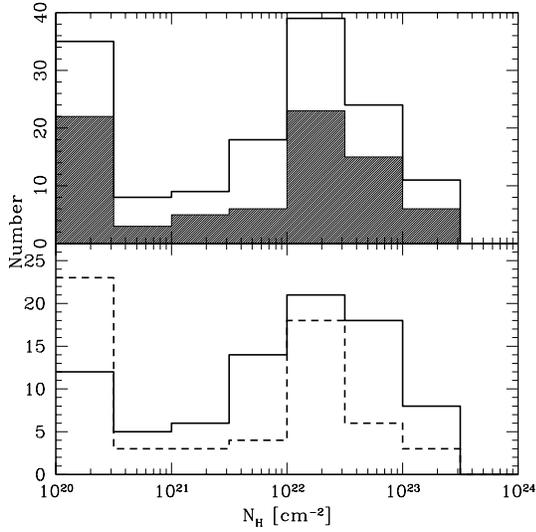}
\end{center}
\caption{{\it Top panel}: Distribution of $N_H$ for all the sources with either spectroscopic or
photometric redshift and more than 80 counts in X-rays. The {\it hatched histogram} shows
the distribution for sources with spectroscopic redshifts only. According to a KS test,
the distributions are entirely consistent with being drawn from the same parent
distribution. {\it Bottom panel}: $N_H$ distribution for sources optically
classified as obscured ({\it solid histogram}) and unobscured ({\it dashed
histogram}). While X-ray obscuration is present in unobscured AGN, most of them have very
small $N_H$ values. Similarly, sources classified as obscured AGN in the optical have in
general higher $N_H$ values. However, in both cases significant scatter is present.}
\label{nh_dist}
\end{figure}

The conversion from observed to intrinsic $N_H$ can thus only be done for sources with a
measured redshift. We have spectroscopic redshifts, given in Table~\ref{cat}, for a total
of 80 sources with fitted $N_H$. However, in order to increase our sample we used the good
quality photometric redshifts derived by the COMBO-17 survey, which provides redshifts
accurate to $\delta$$z$/(1+$z$)$\sim$10\% for sources with $R$$<$24 \citep{wolf04}. This
uncertainty in redshift translates into an uncertainty in the derived intrinsic $N_H$ of
$\delta \log N_H$$\simeq$0.12, smaller than the typical errors in the fitted observed
$N_H$ or the bin size in our $N_H$ distribution (Fig.~\ref{nh_dist}). Hence, it is
acceptable to use the COMBO-17 photometric redshifts in order to convert observed $N_H$
into intrinsic values, thus increasing our sample to 144 sources. To confirm that the use
of COMBO-17 redshifts does not bias our conclusions, we performed a Kolmogorov-Smirnov
(KS) test and found that the hypothesis that both distributions were drawn from the same
parent distribution can only be dismissed at the 0.02\% confidence level.

\section{Discussion}
\label{discussion}

\subsection{X-ray/Optical Classification}

\begin{figure}
\begin{center}
\plotone{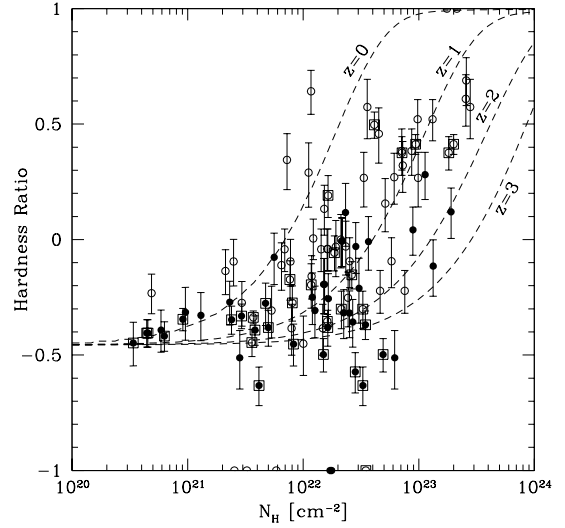}
\end{center}
\caption{Neutral hydrogen column density, $N_H$, as a function of hardness ratio.
Sources optically classified as unobscured AGN are shown by {\it solid circles}, while
obscured AGN are marked by {\it open circles}. Sources with more than 200 X-ray counts
detected are enclosed in squares. The {\it dashed lines} show the relation between $N_H$
and HR for a power-law source with $\Gamma$=1.9 at $z$=0,1,2 and 3. At
$N_H$$\leq$10$^{22.5}$, and in particular at high redshift, the $N_H$ measurements are
very sensitive to small changes in HR. This can explain the presence of optically
unobscured sources with high $N_H$ values.}
\label{nh_hr}
\end{figure}

The bottom panel of Figure~\ref{nh_dist} shows the distribution of intrinsic $N_H$
separately for sources classified as obscured and unobscured AGN based only on optical
spectroscopy. While the average value of $N_H$ is larger for obscured than unobscured
sources, 8$\times$10$^{21}$~cm$^{-2}$ versus 2$\times$10$^{21}$~cm$^{-2}$, the difference
is not very large. Performing a KS test to these two distributions rules out the
hypothesis that they were drawn from the same parent distribution with a 98.5\%
confidence, leaving a small but not completely negligible probability. Since they are both
indications of the X-ray spectral shape, there is a close relationship between the
hardness ratio and derived $N_H$ values. As can be seen in Figure~\ref{nh_hr}, for modest
column densities, $N_H$$\leq$10$^{22.5}$ the $N_H$ values are very sensitive to small
changes in the number of detected soft X-ray photons, in particular for high redshift
sources. This implies that there can be large errors in the derived $N_H$ values, in
particular for unobscured sources at high redshift. The implications of this potential
bias in our analysis will be further discussed below.

\begin{figure}
\begin{center}
\plottwo{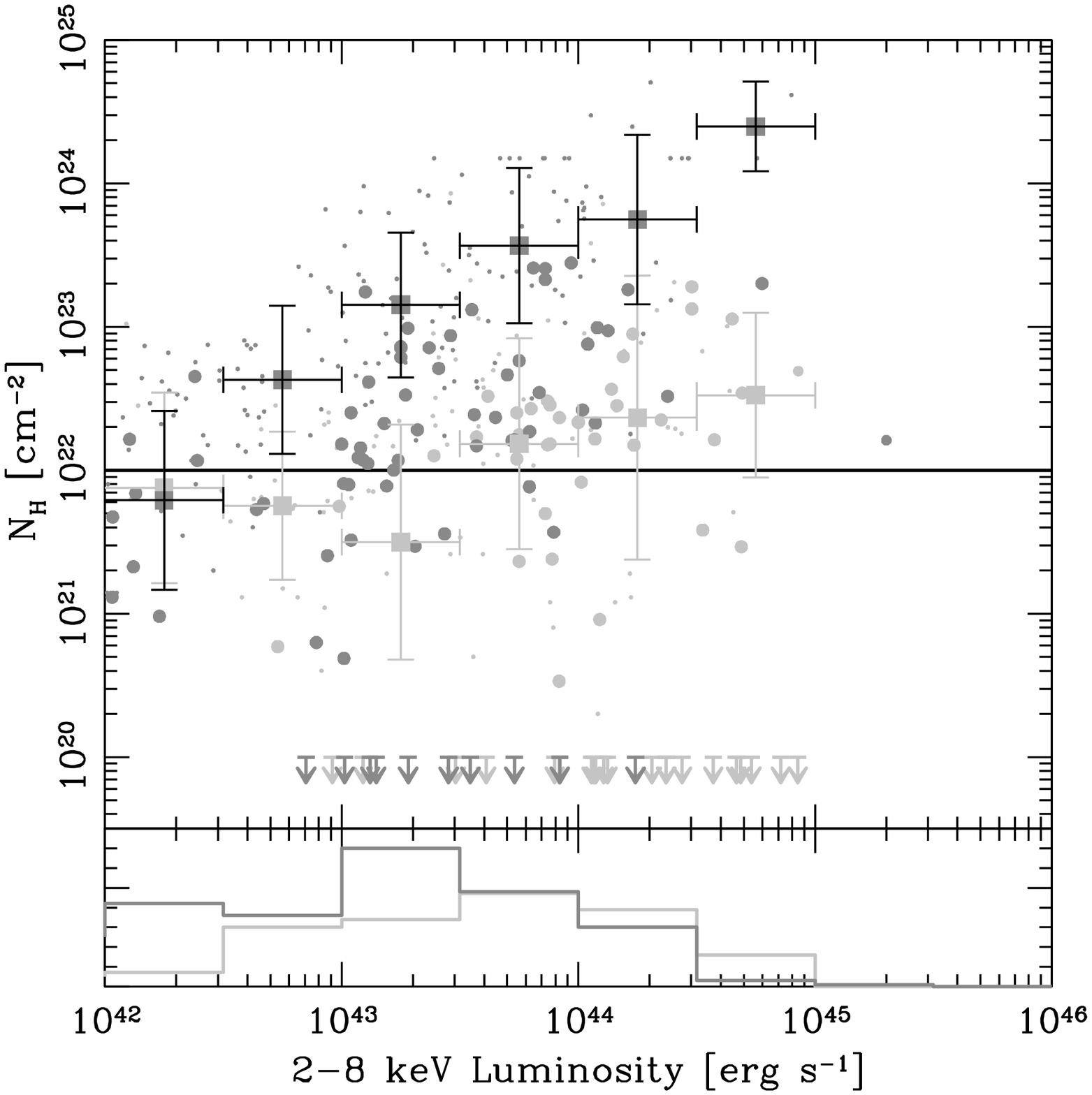}{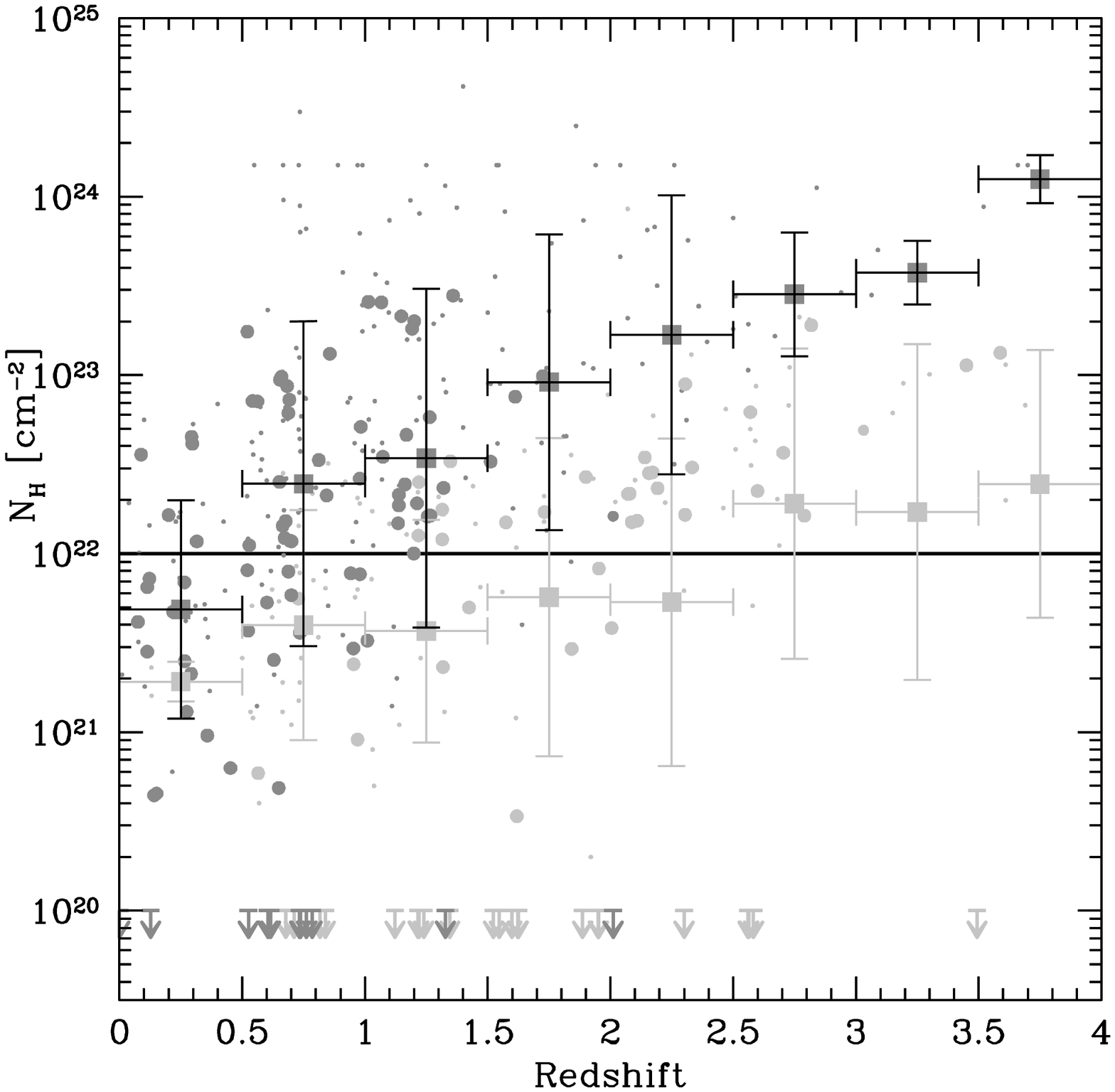}
\end{center}
\caption{{\it Left panel}: $N_H$ as a function of hard X-ray luminosity for sources optically 
classified as unobscured ({\it light gray circles}) and obscured ({\it black}) AGN. The
large circles show the AGN in our sample, while the small circles are sources in the
deeper central region with $N_H$ measured by \citet{tozzi06}. The {\it lower panel} shows
the luminosity distribution for the same sources. Large symbols with errorbars show the
average values in bins of 20 sources. A clear separation in the $N_H$ average can be seen
for obscured and unobscured sources, except at low luminosities, where the effects of
dilution by the host galaxy can hide the broad emission lines used to classify sources
optically. {\it Right panel}: Same as left panel, but showing $N_H$ as a function of
redshift. In this case there is a clear increasing trend in the average values with
redshift. This can be explained by the redshift correction in converting observed into
intrinsic $N_H$ values. Because of this correction, sources optically classified as
unobscured at high redshift, can appear to have large $N_H$ values, as discussed in the
text.}
\label{lum_red_nh}
\end{figure}

As was shown in Figures \ref{hr_lum} and \ref{nh_hr}, the optical and
X-ray classifications do not always agree. In order to further
investigate this effect, in the left panel of Fig.~\ref{lum_red_nh} we
show the intrinsic $N_H$ versus hard X-ray luminosity, used as a
tracer of the intrinsic AGN luminosity. For this figure we also added
335 X-ray sources detected in the 1~Msec. CDF-S observations with
accurate $N_H$ measurements based on spectral fitting presented by
\citet{tozzi06}. Redshifts for those sources, either photometric or
spectroscopic, were compiled by \citet{zheng04}. This figure shows two
interesting effects. At low luminosities,
$L_X$$<$10$^{43}$~erg~s$^{-1}$, a large fraction of the sources
optically classified as obscured AGN have intrinsic $N_H$ values lower
than 10$^{22}$~cm$^{-2}$. This can be explained by the effects of
dilution of the AGN light by the host galaxy (e.g.,
\citealp{moran02,cardamone07}), which lowers the equivalent width of
the broad emission lines because of the relatively high continuum from
the host galaxy, thus making a source to be erroneusly classified as
obscured AGN in the optical. A second effect appears at high
luminosity: sources optically classified as unobscured AGN have
intrinsic $N_H$ values higher than 10$^{22}$~cm$^{-2}$. A possible
explanation for this effect can be found in the right panel of
figure~\ref{lum_red_nh}, where we can see that the sources with
discrepant X-ray/optical classifications are predominantly located at
$z$$>$2.5. Given the strong redshift dependence of the accuracy of
$N_H$ determinations, sources with very small observed $N_H$ values,
which can be in general explained by measurement errors, will have a
high intrinsic $N_H$ values if the source is located at high
redshift. At high redshift, the Chandra observations trace
emission at higher energy in the rest frame, thus making it much
harder to detect the photoelectric absorption cutoff, critical for
measuring $N_H$. This effect was already studied and simulated by
\citet{akylas06}, who found that at high redshift the derived column
densities are systematically overestimated. According to the results
of \citet{akylas06}, $N_H$ measurements are systematically
overestimated by $\sim$50\% at $z$$\sim$2.5 and $\sim$20\% at
$z$$\sim$1.5.

An obvious conclusion of this analysis is that no classification method is perfect and can
be used for all sources. An early attempt to use a combined X-ray/optical classification
was performed by \citet{szokoly04}. In their work, they used a separation between obscured
and unobscured sources at a hardness ratio of -0.2. A similar approach was used by
\citet{hasinger05} to select unobscured AGN from an X-ray selected samples. This threshold
is shown in Figure~\ref{hr_lum} for the sources with spectroscopic identification in the
ECDF-S. A hardness ratio of -0.2 corresponds to an intrinsic $N_H$ of 10$^{22}$~cm$^{-2}$
at $z$=0.5, assuming $\Gamma$=1.9, while sources with this $N_H$ at lower redshift will
have a higher (more positive) value. So, a threshold of -0.2 in HR to separate obscured
and unobscured AGN is appropriate at $z$$<$0.5. However, as is clear from
Figure~\ref{lum_red_nh}, the effects of dilution by the host galaxy are important only for
low luminosity sources, which are predominantly located at low redshift. Hence, we propose
the following classification scheme: For sources at $z$$<$0.5 we will separate obscured
and unobscured sources at HR=-0.2, regardless of their optical properties, while for
sources at higher redshifts we will use only the optical classification scheme, as $N_H$
measurements are biased at high redshift.

\begin{figure}
\begin{center}
\plottwo{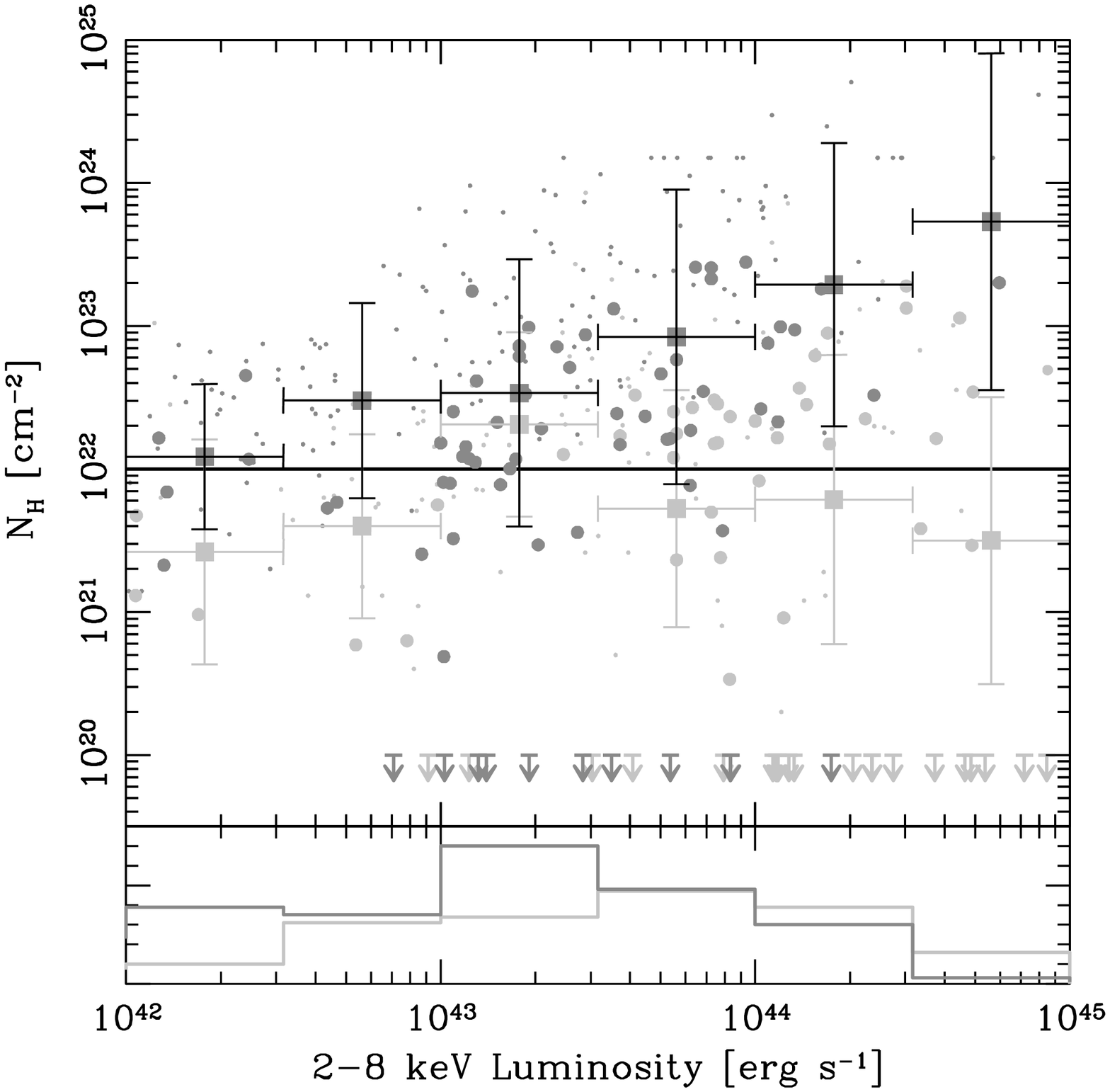}{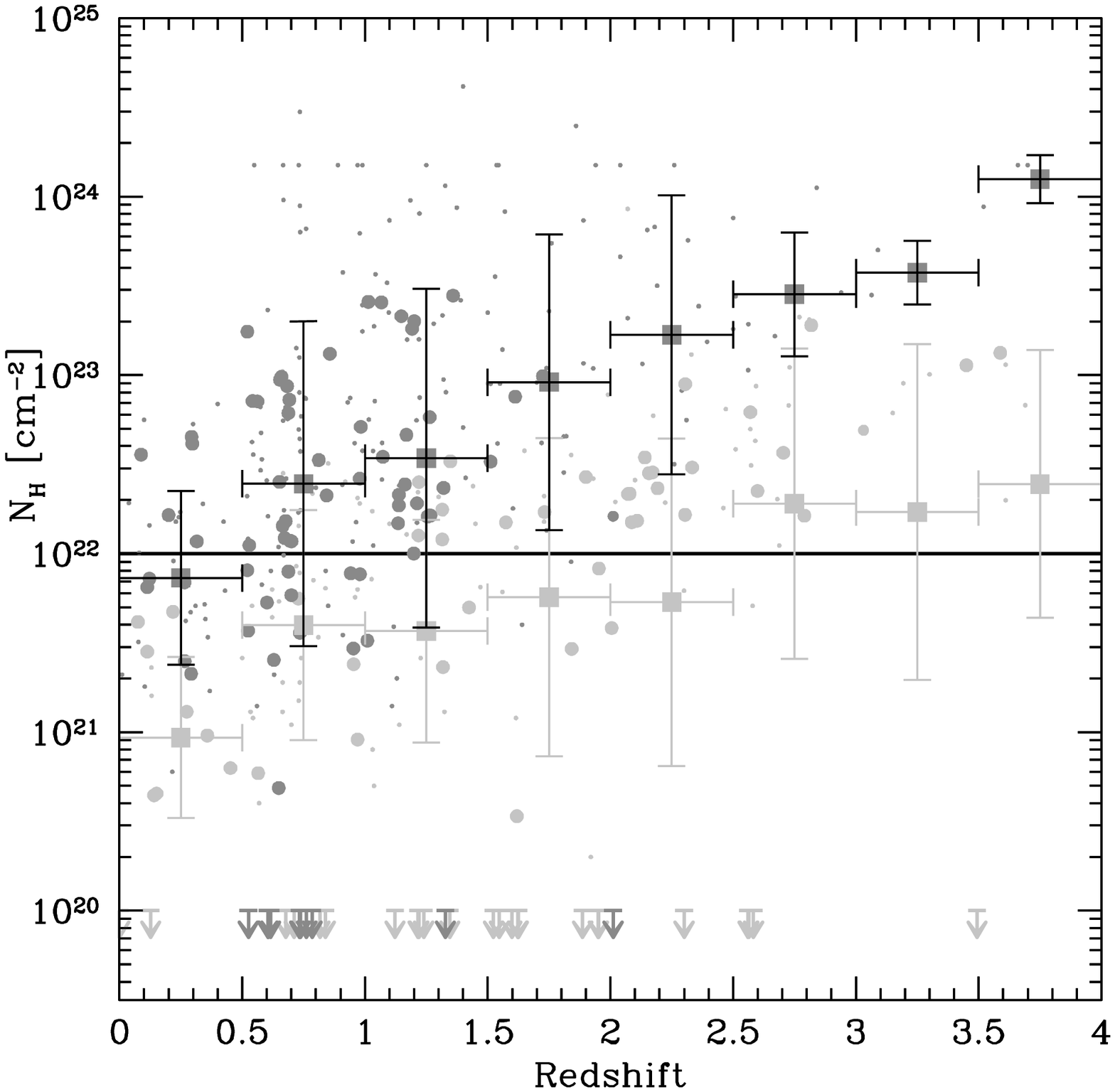}
\end{center}
\caption{Same as Fig.~\ref{lum_red_nh} but using the combined classification scheme described in 
the text. In this case, it can be seen that the discrepancy between the optical and X-ray
classifications for low luminosity sources is solved and our hybrid classification clearly
separate sources in their average $N_H$ values.}
\label{lum_red_nh_v2}
\end{figure}

In Figure~\ref{lum_red_nh_v2} we show the measured $N_H$ values as a
function of luminosity and redshift for the new classification
scheme. As expected, changes are relevant only for sources with low
luminosities at low redshifts. With an optical classification only,
the average $N_H$ for obscured sources at $z$$<$0.5 is
4.8$\times$10$^{21}$~cm$^{-2}$, while using the X-ray classification
it is 7.3$\times$10$^{21}$~cm$^{-2}$. Similarly, in the lowest
luminosity bin, $L_X$$<$3.2$\times$10$^{42}$~erg~s$^{-1}$, obscured
and unobscured AGN are now clearly separated in $N_H$. This separation
is not even larger since heavily-obscured low-luminosity sources are
preferentially missed by the X-ray observations, and thus not included
in our sample.

\subsection{Rest-frame Optical Colors}

As is well known (e.g., \citealp{baldry04,weiner05,cirasuolo07} and
references therein), the distribution of optical colors in normal
galaxies is bimodal, with a large population of blue star-forming
galaxies in the blue cloud separated from the ``red sequence'' of
massive passively-evolving spheroids. As was recently reported (e.g.,
\citealp{barger03,nandra07,georgakakis08,silverman08}), most galaxies
hosting obscured AGN, in which the integrated optical light is
dominated by the host galaxy (e.g., \citealp{treister05a}), lie on the
``green valley'', a transition region located between the red sequence
and the blue cloud. Hence, this can be considered as good evidence for
a link between AGN and galaxy evolution, possibly indicating that AGN
feedback can play a role in regulating star formation
\citep{springel05,schawinski06}; however, this remains controversial,
as discussed by \citet{alonso-herrero08}.

\begin{figure}
\begin{center}
\plotone{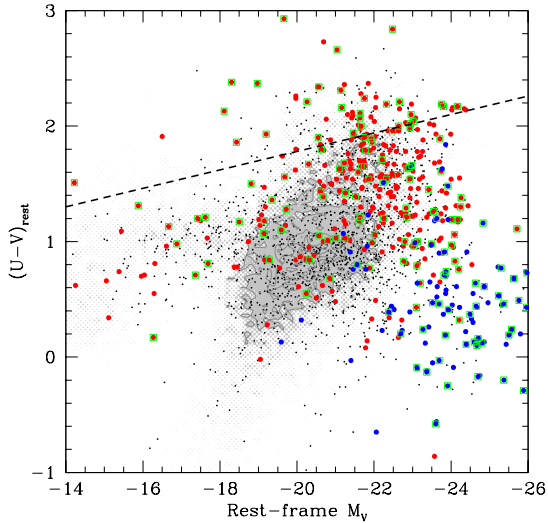}
\end{center}
\caption{Rest frame $U$-$V$ color versus rest-frame $V$-band absolute magnitude for obscured 
({\it red circles}) and unobscured AGN ({\it blue circles}). The {\it green boxes} mark
the location of sources with spectroscopic identifications. Normal galaxies (i.e.,
non-AGN) with spectroscopic redshifts are shown by {\it black circles}, while the gray
background shows the density of normal galaxies with photometric redshifts from
COMBO-17. Contours enclose regions with densities larger than 10, 20 and 30 galaxies per
interval. While the unobscured AGN are in general bluer and have higher luminosities than
normal galaxies, obscured AGN have similar optical luminosities but have redder $U$-$V$
colors and live around the ``green valley'', shown by the {\it dashed line}, as determined
by \citet{bell04} for $z$=1.}
\label{uv_v}
\end{figure}

To compute the rest-frame optical colors for the X-ray sources with
spectroscopic identification on the ECDF-S field we started from the
MUSYC optical catalog of \citet{gawiser06b}. In order to increase the
number of sources in our sample we added the X-ray sources with
reliable photometric redshifts from the COMBO-17 survey, as described
before. We then generated the rest-frame spectral energy distribution
for each X-ray source with measured redshift by interpolating the
observed $UBVRIzJK_s$ photometric data points. The rest-frame $U$ and
$V$ optical magnitudes were then computed by convolving the energy
distribution with the corresponding filter response. The resulting
rest-frame $U$-$V$ colors versus $M_V$, the rest-frame $V$-band
absolute magnitude, are shown in Fig.~\ref{uv_v}. A significant number
of X-ray sources have $U$-$V$$<$1 and $M_V$$<$-22. In general, those
sources were classified as unobscured AGN because of the presence of
broad emission lines. Since the optical continuum of these sources is
dominated by the AGN and not the host galaxy, we exclude them from
further analysis. However, as it was found by \citet{schawinski08},
once the emission from the central source is removed, the host
galaxies of unobscured AGN have optical colors similar to those of
unobscured AGN. As can be seen by comparing with a population of
normal galaxies obtained on the same field from the COMBO-17 survey,
the X-ray sources classified as obscured AGN have in general redder
optical colors than normal galaxies and lie outside of the main blue
cloud.

\begin{figure}
\begin{center}
\plotone{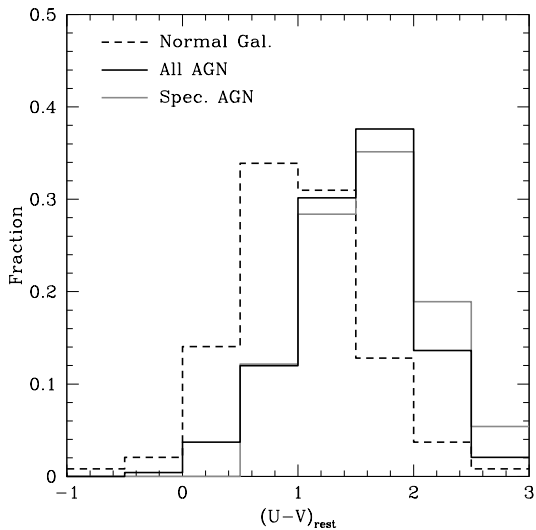}
\end{center}
\caption{Normalized distribution of $U$-$V$ values for normal galaxies ({\it dashed line}), 
AGN ({\it black solid line}) and AGN with spectroscopic identification ({\it gray line})
for sources with -23$<$$M_V$$<$-18. The sample of normal galaxies was selected randomly to
match the AGN redshift distribution. The separation between normal galaxies and AGN in
their $U$-$V$ colors can be seen clearly. In this range of optical luminosities, most of
the AGN are obscured, hence the integrated emission is dominated by the galaxy, thus
indicating that in general AGN host galaxies have redder color than non-active galaxies.}
\label{uv_dist}
\end{figure}

The excess of red sources in the AGN population can be easily seen in
Fig.~\ref{uv_dist}, where the distributions of rest-frame $U$-$V$
colors for X-ray emitting sources and normal galaxies are
presented. In order to limit the influence of differences in the
sample selection, the comparison sample was obtained randomly from the
ECDF-S inactive galaxies to match the AGN redshift
distribution. Performing a KS test we found that the probability that
these two distributions were drawn from the same parent distribution
is negligible. Contrarily, a KS test comparing the distribution of the
X-ray sources with spectroscopic and photometric redshift returned a
relatively high probability of $\sim$20\%, indicating that the use of
photometric redshifts does not significantly bias our results. The
fact that results of the KS test are not even higher can be explained
since in general sources with spectroscopic redshifts have brighter
optical counterpart, and thus slightly different average $U$-$V$
colors. However, this is a minor effect that will not affect our
conclusions.

Given that the comparison sample was selected to have the same
absolute optical magnitude and redshift distributions as the AGN host
galaxies, it is unlikely that the differences in optical colors are
due to the sample selection. A similar analysis was performed by
\citet{georgakakis08} using the AGN detected in the AEGIS
survey. Studying the morphology of AGN host galaxies, they found only
a small fraction of major mergers, but evidence of minor interactions
in a large number of sources. Both \citet{georgakakis08} and
\citet{schawinski08} concluded that AGN activity should outlive the
truncation of star formation, in order to explain the observed red optical
colors of AGN host galaxies.

\subsection{Obscured AGN Fraction}

\begin{figure}
\begin{center}
\plotone{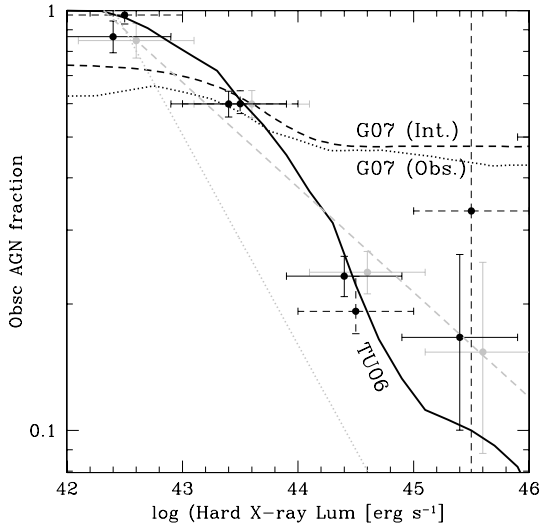}
\end{center}
\caption{Ratio of obscured to total AGN as a function of hard X-ray luminosity. The 
black circles with {\it dashed error bars} show the obscured AGN fraction from the ECDF-S
alone, while the {\it gray circles} show the results obtained using the sample of
\citet{treister06b}. Black circles with {\it solid error bars} show the fraction of
obscured AGN for the combined sample of \citet{treister06b} and the ECDF-S sources. The
dependence used by the \citet{gilli07} AGN population synthesis model for the intrinsic
and observed fractions of obscured AGN are shown by the {\it dashed} and {\it dotted}
lines. The dependence used in the model presented by \citet{treister06b} is shown by the
{\it solid line}. While a good agreement is found between the observations and the
expectations of \citet{treister06b}, the \citet{gilli07} model predicts a relatively large
fraction of obscured AGN, $\sim$50\%, at high luminosities, while the observed value is
only $\sim$20\%. The {\it dashed gray line} shows the expected dependence for a
radiation-limited torus as described by \citet{hoenig07}, while the {\it dotted gray line}
shows the expectation for the original receding torus of \citet{lawrence91}, both
normalized to the observed value in the 10$^{42-43}$~erg~s$^{-1}$ bin.}
\label{obs_frac_lum}
\end{figure}

The fraction of obscured AGN and its possible dependence on parameters
like redshift or luminosity is very important for AGN population
studies, in particular given its direct relation with our
understanding of the cosmic X-ray Background. A dependence of the
obscured fraction on luminosity was previously found for X-ray
selected samples by
\citet{ueda03}, \citet{steffen03}, \citet{lafranca05} and \citet{treister05a} among others. 
A physical explanation, proposed by \citet{lawrence91} and updated by
\citet{simpson05}, is the so-called ``receding torus''. More recently, \citet{treister08}
found a dependence of the ratio of mid-IR to bolometric flux on luminosity for unobscured
AGN consistent with an increase in the opening angle with luminosity. Similarly,
\citet{akylas08} proposed that the observed dependence of the obscured fraction on
luminosity could be explained by the effects of photo-ionization on the X-ray obscuring
matter, while \citet{hoenig07} argues that the effects of the Eddington limit on a clumpy
torus can explain the observed luminosity dependence of the obscured AGN fraction. The
large sample of X-ray sources in the ECDF-S can be used to further explore this luminosity
dependence of the obscured AGN fraction. In figure~\ref{obs_frac_lum} we can see that
using the ECDF-S sample alone the fraction of obscured AGN decreases from $\sim$90\% at
$L_X$$<$10$^{43}$~erg~s$^{-1}$ to $\sim$20\% at $L_X$=10$^{45}$~erg~s$^{-1}$, using the
``hybrid'' classification scheme described in \S 4.1.

In order to increase the significance of this result, we added the results from the ECDF-S
to the compilation presented by \citet{treister06b}, which combined the results from seven
X-ray surveys ranging from wide area shallow surveys to the Chandra deep fields. The total
sample includes now 2814 X-ray sources, 1377 (49\%) of them with spectroscopic
identification. In Figure~\ref{obs_frac_lum} we present the resulting obscured AGN
fraction as a function of luminosity for the total sample. The obscured fraction ranges
from 80$\pm$7\% at $L_X$$<$10$^{43}$~erg~s$^{-1}$ to 16$\pm$8\% for
$L_X$$>$10$^{45}$~erg~s$^{-1}$. In addition, Figure~\ref{obs_frac_lum} shows the predicted
luminosity dependence incorporated in the \citet{treister05b} AGN population synthesis
model, as updated by \citet{treister06b} to include a redshift dependence, and for
comparison the luminosity dependence used in the models of \citet{gilli07}. While at low
luminosities both models agree well with the observations, for luminosities higher than
$L_X$$\simeq$10$^{44}$~erg~s$^{-1}$ the \citet{gilli07} models predict a fraction of
obscured AGN of $\sim$50\%, significantly higher than the observed value. This discrepancy
has important consequences for the modeling of the X-ray background using AGN, as the
largest contribution comes from sources at roughly these luminosities (e.g.,
\citealp{treister05b}). The larger fraction assumed by \citet{gilli07} is more relevant at
lower energies, where the effects of absorption are larger. For example, the XRB intensity
at E$<$10~keV in the \citet{gilli07} model is $\sim$40\% lower than the
\citet{treister05b} calculation, while it is now clear from recent Chandra and XMM
observations \citep{deluca04,hickox06} that a larger XRB intensity at low energies is more
appropriate.

In addition, in Figure~\ref{obs_frac_lum} we compare the observed dependence of the
fraction of obscured sources on luminosity with the expectations for different geometrical
parameters of the obscuring material. If the height of the torus is roughly independent of
luminosity, the change in covering fraction is due to a change in inner radius (the
original ``receding torus'' model), hence a rough $L^{-1/2}$ dependence for the contrast
should be expected \citep{barvainis87,lawrence91}. If the effects of radiation pressure
are incorporated, in the case of a clumpy torus, \citet{hoenig07} derive a $L^{-1/4}$
dependence for the contrast. As can be seen in Figure~\ref{obs_frac_lum}, a $L^{-1/2}$
dependence is too steep compared to observed data. This implies that the height of the
obscuring material cannot be independent of the source luminosity and provides evidence for
a radiation-limited structure, as the one suggested by \citet{hoenig07}.

\begin{figure}
\begin{center}
\plotone{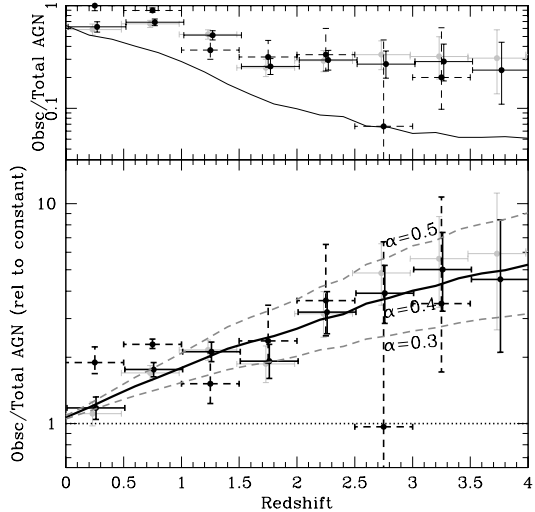}
\end{center}
\caption{Fraction of obscured AGN as a function of redshift. {\it Upper panel}: direct measurements 
using the sources on the ECDF-S field only ({\it black circles with dashed error bars}),
from the \citet{treister06b} sample ({\it gray circles}) and combining both samples ({\it
black circles with solid error bars}). The expected observed fraction for an intrinsic
fraction of 3:1 obscured to unobscured AGN including optical and X-ray selection effects
is shown by the {\it black solid line}. As can be seen, while the observed fraction of
obscured AGN declines toward higher redshifts, if the X-ray and optical selection effects
and the luminosity dependence of the obscured AGN fraction are taken into account this
decline should be even stronger. {\it Bottom panel}: Inferred fraction of obscured AGN
relative to an intrinsically constant fraction after correcting for selection effect and
including the luminosity dependence of the obscured AGN fraction. Symbols are the same as
for the upper panel. The corrected fraction of obscured AGN increases with redshift
following a power-law of the form (1+$z$)$^\alpha$ with $\alpha$=0.4$\pm$0.1, consistent
with the results found by \citet{treister06b}.}
\label{obs_frac_red}
\end{figure}

The dependence of the fraction of obscured AGN on redshift is more controversial. While
some studies (e.g., \citealp{lafranca05,ballantyne06,treister06b,dellaceca08}) found a
small increase in the fraction of obscured AGN at higher redshifts, other results suggest
that this fraction is constant (e.g., \citealp{ueda03}, \citealp{akylas06}). The upper panel
of Fig.~\ref{obs_frac_red} shows the observed fraction of obscured AGN as a function of
redshift for the sources in the ECDF-S. This fraction is high, $\sim$90\%, at $z$$<$1,
while at higher redshifts it decreases to $\sim$30-40\%. As for the luminosity dependence,
in order to increase the significance of our results we added the ECDF-S sources to the
sample of \citet{treister06b}. The results for the large sample are consistent with those
from the ECDF-S alone; namely, at low redshift there is a large fraction of obscured
sources with a steep decline at $z$$\sim$1. This decline in the observed fraction of
obscured sources can be easily explained by a simple observational fact: in order to be
included in this sample a source needs to have a measured redshift from optical
spectroscopy. For obscured AGN, the optical light is dominated by the host galaxy (e.g.,
\citealp{barger05,treister05a}), which becomes too faint for spectroscopy with 8 meter
class telescopes at $z$$\sim$1. In order to quantify this effect, we use the ratio of
identified to total X-ray sources in a given optical magnitude bin (upper panel,
fig.~\ref{r_dist}). This magnitude-dependent ratio is used to calculate the expected
obscured AGN fraction including the X-ray and optical flux limits and the luminosity
dependence of the obscured fraction. This calculation is done as described in detail by
\citet{treister06b}. Briefly, we used the AGN population synthesis of \citet{treister05b},
together with the library of AGN spectra described by \citet{treister04}, and calculated
the effective area of the survey as a function of both X-ray flux and optical magnitude,
taking into account the spectroscopic incompleteness at each optical flux. This procedure
outputs a expected number of {\it observed} obscured and unobscured AGN as a function of
redshift, for a non-evolving obscured AGN fraction.

As can be seen in the upper panel of Figure~\ref{obs_frac_red}, the expected decrease is
actually steeper than what is observed. This implies that the inferred fraction of
obscured AGN in the ECDF-S should increase with redshift, once the selection effects in
the sample are accounted for. This dependence is consistent with the results of
\citet{treister06b}, who found that this increase can be represented as a power-law of the
form (1+$z$)$^\alpha$, with $\alpha$=0.4$\pm$0.1. These results are roughly independent of
the method used to classify AGN; we obtained the same dependence using both our mixed
X-ray/optical classification scheme and one based completely on optical spectroscopy.

\subsection{Number Density and Evolution of AGN}

Using our combined sample of X-ray selected AGN we can also compute the AGN spatial
density as a function of redshift, and compare with the expectations from existing AGN
luminosity functions based on smaller samples. In order to calculate the AGN spatial
density, we started from our collected sample of 1377 sources described above. We then
separated the sources in low ($L_X$=10$^{41.5-43}$~erg~s$^{-1}$), medium
($L_X$=10$^{43-44.5}$~erg~s$^{-1}$) and high luminosity
($L_X$=10$^{44.5-48}$~erg~s$^{-1}$) bins. For each luminosity class we binned the sample
in redshift so that each bin has at least 50 sources (20 for the highest luminosity
sources). Then, the spatial density of each bin was calculated by summing the values of
$V_c^{-1}$ for each source in the bin, where $V_c$ is the total comoving volume per unit
area in that bin multiplied by the area covered by our super-sample at the X-ray flux of
the source. In addition, upper limits for the spatial density on each bin were calculated
by multiplying the values of $V_c$ by the fraction of spectroscopically identified sources
at the optical magnitude of the AGN, as described in \citet{treister06b} for the general
sample and in the upper panel of Fig.~\ref{r_dist} for the ECDF-S.

\begin{figure}
\begin{center}
\plotone{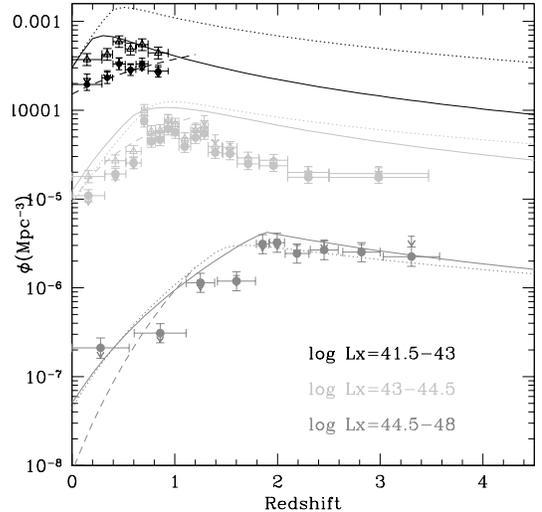}
\end{center}
\caption{Space density of AGN as a function of redshift derived from the sources in 
the ECDF-S. The {\it filled circles} show the observed density versus $z$ in three luminosity bins,
$\log L_X$=41.5-43, 43-44.5 and 44.5-48~erg~s$^{-1}$. Arrows show the effects of
correcting these values for optical incompleteness effects, while the {\it empty triangles} show the results using
the $N^\textnormal{obs}/N^\textnormal{mdl}$ method to correct for incompleteness in the X-ray data. The lines show the AGN space
density derived using the luminosity functions of Ueda et al. (2003; {\it solid lines}),
Barger et al. (2005; {\it dashed lines}) and La Franca et al. (2005; {\it dotted
lines}). The \citet{barger05} luminosity function is only defined for $z$$<$1.2. While in
the highest luminosity bin all the studied luminosity functions agree very well with the
observations, at intermediate luminosities a good agreement is found only at $z$$<$1. 
In the lowest luminosity bin, the effects of incompleteness are more important, and hence only
after correcting for them a good agreement with the luminosity function of \citet{ueda03} is found.
The \citet{barger05} luminosity function was not corrected for incompleteness.}
\label{dens_red}
\end{figure}

In Fig.~\ref{dens_red} we show the observed AGN spatial density as a function of redshift
for low, medium and high luminosity sources compared with the expected density obtained
from integrating the hard X-ray luminosity functions presented by \citet{ueda03},
\citet{barger05} and \citet{lafranca05}. While the X-ray sources in these studies were selected
in similar ways, the samples have different numbers of sources (247 AGN in \citealp{ueda03},
746 in \citealp{barger05}, and 508 in \citealp{lafranca05}) with the emphasis set at different
flux levels (mostly bright sources in \citealp{ueda03}, moderate fluxes in
\citealp{lafranca05} and faint sources in \citealp{barger05}). In addition, the modeling of
the luminosity function is also different in these papers. While \citet{barger05} assumed
pure luminosity evolution, both \citet{ueda03} and \citet{lafranca05} found better results
using a luminosity-dependent density evolution. As can be seen in Fig.~\ref{dens_red}, at
high and moderate luminosities all the luminosity functions studied here agree well with
the observations at low redshifts, while at high redshifts the expectations from both the
\citet{ueda03} and \citet{lafranca05} works are significantly above the observed values
for intermediate luminosity sources. 

This discrepancy can most likely be explained by the effects of
incompleteness in the X-ray data, which are particularly important for
the lower luminosity sources. As it was concluded by e.g.,
\citet{lafranca97}, the $V_c^{-1}$ method used here is particularly
sensitive to the effects of incompleteness. The work of \citet{ueda03}
and \citet{lafranca05} both used the
$N^\textnormal{obs}$/$N^\textnormal{mdl}$ method, which attempts to
account for incompleteness by using the expected number of sources in
a given luminosity and redshift bin. One obvious caveat of this method
is that the results are model dependent. We attempt to correct for
incompleteness in the X-ray data, in the two lowest luminosity bins,
where these effects are more important. In order to do that, we use
the model described in the previous section in order to calculate the
fraction of sources missed by the X-ray selection in a given
luminosity and redshift bin due to the effects of obscuration. For the
lowest luminosity sources, this correction is roughly $\sim$2$\times$,
while at higher luminosities it is about 30-40\%. We did not attempt
to correct the highest luminosity bin, since the effects of absorption
are negligible for these sources. As can be seen in
Fig.~\ref{dens_red}, after the correction for incompleteness is
applied a good agreement with the predicted spatial density using the
luminosity function of \citet{ueda03} is found. The remaining discrepancy with
the work of \citet{lafranca05} could be explained because in this case
even highly obscured Compton-thick AGN up to
$N_H$=10$^{26}$~cm$^{-2}$ are included, while in the case of \citet{ueda03} and in
our work only mildly Compton-thick AGN with
$N_H$$<$10$^{25}$~cm$^{-2}$ are considered. Finally, it is worth
mentioning that in the luminosity function of \citet{barger05} not
attempts were made to correct for incompleteness in the X-ray
selection. Hence, a good agreement is found only for the uncorrected
data points.

\subsection{Infrared (IR) to X-ray ratio}

Emission at X-ray wavelengths, in particular in the hard band, is often used as a tracer
for the direct AGN output mainly from Compton-scattered accretion disk photons, while the
luminosity at longer wavelengths, in particular in the mid-far infrared, is associated
with the re-radiation of the energy absorbed by the surrounding gas and dust (e.g.,
\citealp{pier93}). There is currently a strong debate about the geometry and
characteristics of this surrounding dust, in particular whether it has a smooth (e.g.,
\citealp{pier92}) or clumpy (e.g., \citealp{krolik88}) distribution. The ratio of IR to
X-ray luminosity can be used to distinguish between these two distributions and to
constrain the dust geometry \citep{lutz04}. For example, smooth torus models predict large
differences in the IR to X-ray ratio for obscured and unobscured sources, due to the
significant effects of self-absorption. Contrarily, radiation transfer models of clumpy
dust torii predict very small or no differences between the IR to X-ray ratio for obscured
and unobscured AGN.

The observational evidence remains controversial. The work of \citet{horst06} reports that
no differences were found in the mid-IR (measured at a fiducial rest-frame wavelength of
12.3~$\mu$m) to X-ray ratio for a sample of 17 nearby AGN observed with the VLT-VISIR,
which provides a relatively high angular resolution of $\sim$0.35$''$. In apparent
contradiction, \citet{ramos-almeida07} found a slightly smaller value for the X-ray to
nuclear mid-IR (at 6.75~$\mu$m) ratio from a sample of 57 AGN. For the latter, the
observations were carried out using the ISOCAM camera onboard ISO, with a more limited
angular resolution of $\sim$4$''$. \citet{horst08} argue that precisely this difference in
angular resolution explains the discrepant results. Using their high spatial resolution
images, they claim that the contribution from star formation, unresolved in the ISOCAM
observations, can account for the observed differences between obscured and unobscured
AGN. On the other hand, the relatively small sample of \citet{horst06} traces larger
intrinsic luminosities for unobscured sources compared to the obscured sample, thus
contributing to explain why no difference in the IR to X-ray ratio was found. Given the
large scatter in the IR to X-ray ratio reported by both groups, a large sample of sources
is required to reach statistically significant conclusions.

\begin{figure}
\begin{center}
\plotone{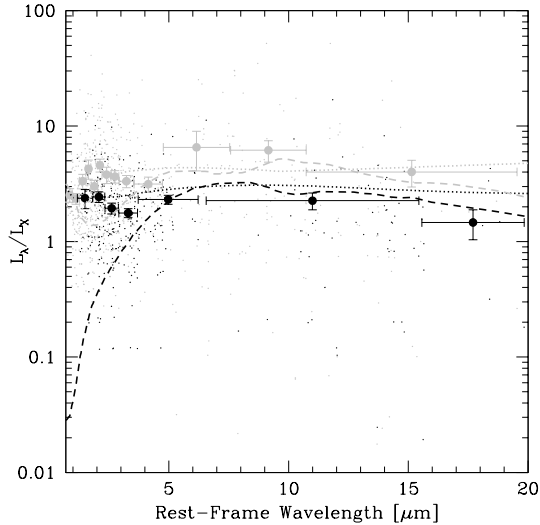}
\end{center}
\caption{Monochromatic luminosity normalized by the hard X-ray luminosity as a function of
rest-frame wavelength for obscured ({\it small black circles}) and unobscured ({\it
small light gray circles}) AGN. For each source the four Spitzer IRAC and the
MIPS-24~$\mu$m photometric data points were used. The {\it large circles} with error bars
show the average values for each AGN type, in bins that include 40 sources. The dotted
lines show the composite spectra for obscured and unobscured quasars, as compiled by
\citet{polletta07}, while the dashed lines show examples of expected IR spectrum from
the clumpy torus models of \citet{nenkova08}, as described in the
text. A clear separation between the average values for obscured and
unobscured sources is observed at wavelengths $\sim$5-15~$\mu$m, which
can be explained by the effects of self-absorption by the surrounding
material in obscured sources.}
\label{Lir_x_wave}
\end{figure}

In Figure~\ref{Lir_x_wave} we present the values of the ratio of $L_\lambda$ to $L_X$ as a
function of $\lambda$, the rest-frame wavelength. These values were computed using both
the IRAC and MIPS fluxes, for the sources in our sample with Spitzer detections and $N_H$
measurements. We also added the sources in the central region with $N_H$ measured from the
X-ray spectrum by \citet{tozzi06}. In order to correct the hard X-ray (2--8~keV)
luminosities, $L_X$, for the effects of absorption we used the photoelectric absorption
cross sections derived by \citet{morrison83} and the $N_H$ values measured from X-ray
spectral fitting as described above. Given that the effects of X-ray absorption in sources
classified as unobscured are typically very small, and those values can be significantly
affected by measurement errors, we only corrected the X-ray luminosities of the obscured
sources. If no correction for absorption is done to the X-ray luminosities of the obscured
sources, our conclusions are unchanged. As can be seen in Fig~\ref{Lir_x_wave}, sources
classified as obscured and unobscured AGN have significantly different average values of
$L_\lambda$/$L_X$, and this separation depends on wavelength. At the shortest wavelengths,
$\lambda$$\sim$1~$\mu$m, the separation in this ratio is rather small, which can be
explained by the contribution of the stellar light of the host galaxy to the integrated
emission. This effect remains visible until $\lambda$$\sim$2~$\mu$m, where the stellar
light starts to fade and emission from the host dust in the inner region of the
surrounding material in the AGN begins to dominate. This emission is highly affected by
self-absorption, as most torus models predict (see, e.g.,
\citealp{pier93,nenkova02,hoenig06}), thus explaining why at these wavelengths unobscured
sources have significantly higher values of $L_\lambda$/$L_X$. At longer wavelengths,
$\lambda$$\gtrsim$10~$\mu$m, the contrast between obscured and unobscured sources is
reduced again, since the optical depth is reduced at longer wavelengths. Thus, it is
expected that at $\lambda$$\sim$30-40~$\mu$m the IR emission should become isotropic
again.

In order to compare with the observed averages for nearby Seyfert
galaxies, in Figure~\ref{Lir_x_wave} we also present the composite IR
spectrum for local sources classified as Seyfert 1 (unobscured) and
Seyfert 2 (obscured), as compiled by \citet{polletta07}, normalized at
a fiducial wavelength of 100~$\mu$m, where the IR re-emission is
expected to be fully isotropic. It is remarkable that both composite
spectra agree well with our average values, thus indicating that the
sources in our sample, most of them at $z$=0.5-1 have very similar IR
spectra to local active galaxies and higher luminosity sources. In
addition, the results presented in Figure~\ref{Lir_x_wave} allow to
explain why \citet{ramos-almeida07} and \citet{horst08} reach
apparently discrepant conclusions. While the work of
\citet{ramos-almeida07} was based in observations at $\sim$6~$\mu$m,
where the contrast between obscured and unobscured sources is nearly
maximal, \citet{horst08} used observations of a limited sample of
sources at $\sim$12~$\mu$m, where the contrast is smaller than at
shorter wavelengths. Thus, it is possible that the results of
\citet{horst08} are dominated by the intrinsic scatter in this
relations, in particular given the low number of sources in their
sample.

In Figure~\ref{Lir_x_wave} we also compare the observed infrared
luminosities with the predictions from torus models recently presented
by \citet{nenkova08}. These models assume a clumpy distribution for
the obscuring material.  While establishing the physical parameters of
the dust surrounding the central region is a difficult task, which is
beyond the scope of this paper, in Fig.~\ref{Lir_x_wave} we show
examples of the expected IR spectrum for obscured and unobscured
sources. The assumed parameters are described by \citet{nenkova08} in
the caption of their Figure~4, for a r$^{-3}$ radial distribution of
clouds. Following the basic assumption of the original AGN unification
paradigm, the only difference between the obscured and unobscured
model is the viewing angle. As can be seen, a decent agreement is
found between the model spectrum and observations, in particular at
wavelengths longer than $\sim$5~$\mu$m. At shorter wavelengths, the
additional contribution from the AGN host galaxy (not included in the
model spectrum) starts to be significant, in particular for the
obscured sources.

\begin{figure}
\begin{center}
\plotone{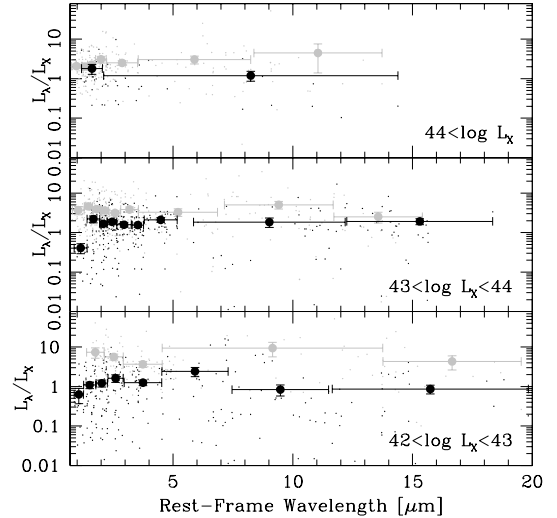}
\end{center}
\caption{Same as Fig.~\ref{Lir_x_wave} but separating the sample in three luminosity
bins: $\log L_X$=42-43, 43-44 and $>$44~erg~s$^{-1}$. The largest separation between
obscured and unobscured sources is found at the lowest luminosities, while for more
luminous sources this difference is smaller and almost negligible. One way to interpret
this result is including a dependence of the opening angle on luminosity, as found by
\citet{treister08}, such that the lower luminosity sources have larger amounts of
self-absorption due to a smaller opening angle.}
\label{Lir_x_wave_lum}
\end{figure}

Taking advantage of the large number of sources in our sample, in
Figure~\ref{Lir_x_wave_lum} we show $L_\lambda$/$L_X$ as a function of
wavelength for sources separated in three luminosity bins:
$L_X$$<$10$^{43}$~erg~s$^{-1}$,
10$^{43}$~erg~s$^{-1}$$<$$L_X$$<$10$^{44}$~erg~s$^{-1}$ and
$L_X$$>$10$^{44}$~erg~s$^{-1}$. A clear conclusion from this figure is
that the difference in the average ratio between obscured and
unobscured sources is largest at the lowest luminosity bin, while for
higher luminosity sources the $L_\lambda$/$L_X$ ratios become very
similar. In their study of the mid-IR properties of nearby AGN,
\citet{polletta07} found somewhat consistent results. According to
their Figure 9, the ratio $L_\textnormal{IR}$ (at rest-frame 6~$\mu$m)
to $L_X$ for obscured sources, obtained combining their AGN2 and SF
classes, is significantly lower than the ratio for unobscured AGN
(their AGN1 class). In our interpretation of
Figure~\ref{Lir_x_wave_lum}, this increasing contrast for lower
luminosity sources can be understood in terms of the luminosity
dependence of the geometrical parameters of the absorbing region, in
particular the opening angle, as described by \citet{treister08}.

\section{Conclusions}

We presented here the first results from our optical spectroscopy program, aimed to
provide identifications and redshifts for the X-ray sources detected by Chandra in the
ECDF-S field. As part of this work we targeted 339 X-ray sources from the catalog of
\citet{virani06} and obtained redshifts and identifications for 160 sources. We also added
26 sources with identifications obtained from the literature. While most of the sources
with broad emission lines are found at $z$$>$1, the mean redshift for the sources in our
sample is 0.75. In order to separate obscured and unobscured AGN we adopted a mixed
scheme, combining the optical spectroscopy information with the indication of the X-ray
spectrum given by the hardness ratio. This hybrid scheme agrees very well in most cases
with the amount of neutral Hydrogen column density measured in the X-ray spectrum for the
brightest sources.

The optical colors of the obscured and/or low luminosity AGN are
dominated by the host galaxy, and are on average redder than
non-active galaxies, as was found previously by other authors. We also
focused on the fraction of obscured AGN and its possible dependence on
luminosity and redshift. For the latter, we confirmed at higher
significance the results of \citet{treister06b}, indicating that after
correcting for selection effects the fraction of obscured AGN
increases with redshift. Regarding a possible luminosity dependence,
we confirmed the functional form derived by \citet{treister05b} and
found that the population synthesis model of \citet{gilli07}
significantly overestimates the fraction of obscured AGN at high
luminosities. We also computed the AGN spatial density as a function
of redshift and found results in agreement with the expected values
using the luminosity function of \citet{barger05}, but significant
discrepancies with the luminosity functions of \citet{ueda03} and
\citet{lafranca05}, which can however by most likely explained by the
effects of incompleteness in our X-ray-selected sample.

Taking advantage of the deep multiwavelength data available in the ECDF-S we studied the
infrared properties of the X-ray selected AGN with spectroscopic identifications. We found
a significant difference in the fraction of bolometric light emitted at mid-IR wavelengths
by obscured and unobscured AGN. These differences can be explained by the effects of dust
self-absorption, for which the maximum contrast should be at
$\sim$5-15~$\mu$m. Furthermore, by separating our sample into luminosity bins we found
that the contrast is larger for the lower luminosity sources. One possible interpretation
for this result is that the opening angle is larger for high luminosity sources, such that
the effects of self absorption are less important.

\acknowledgements

We are grateful for the help and support of the staff at Las Campanas
and Paranal observatories, for very useful discussions with Mark
Gieles and Fabio La Franca. We thank Tom Aldcroft for adapting Yaxx to our
specific needs. Support for the work of ET was provided by the
National Aeronautics and Space Administration through Chandra
Postdoctoral Fellowship Award Number PF8-90055 issued by the Chandra
X-ray Observatory Center, which is operated by the Smithsonian
Astrophysical Observatory for and on behalf of the National
Aeronautics Space Administration under contract NAS8-03060. SV, CMU
and CNC acknowledge support from NSF grant \#AST0407295 and Spitzer
JPL Grant \#RSA1288440. SV acknowledges support from a graduate
research scholarship awarded by the Natural Science and Engineering
Research Council of Canada (NSERC), a Grant-in-Aid of Research from
the National Academy of Sciences (administered by Sigma Xi, The
Scientific Research Society), and from NASA/INTEGRAL grant
NNG05GM79G. This work is based on observations made with the 6.5 m
Magellan telescopes, a collaboration between the Observatories of the
Carnegie Institution of Washington, University of Arizona, Harvard
University, University of Michigan, and Massachusetts Institute of
Technology, and at Cerro Tololo Inter-American Observatory, a division
of the National Optical Astronomy Observatories, which is operated by
the Association of Universities for Research in Astronomy, Inc. under
cooperative agreement with the National Science Foundation. This research has made 
use of NASA's Astrophysics Data System.



\begin{thebibliography}{87}
\expandafter\ifx\csname natexlab\endcsname\relax\def\natexlab#1{#1}\fi

\bibitem[{{Adami} {et~al.}(2005)}]{adami05}
{Adami}, C. {et~al.} 2005, \aap, 443, 805

\bibitem[{{Akylas} \& {Georgantopoulos}(2008)}]{akylas08}
{Akylas}, A. \& {Georgantopoulos}, I. 2008, \aap, 479, 735

\bibitem[{{Akylas} {et~al.}(2006){Akylas}, {Georgantopoulos}, {Georgakakis},
  {Kitsionas}, \& {Hatziminaoglou}}]{akylas06}
{Akylas}, A., {Georgantopoulos}, I., {Georgakakis}, A., {Kitsionas}, S., \&
  {Hatziminaoglou}, E. 2006, \aap, 459, 693

\bibitem[{{Alonso-Herrero} {et~al.}(2008){Alonso-Herrero},
  {P{\'e}rez-Gonz{\'a}lez}, {Rieke}, {Alexander}, {Rigby}, {Papovich},
  {Donley}, \& {Rigopoulou}}]{alonso-herrero08}
{Alonso-Herrero}, A., {P{\'e}rez-Gonz{\'a}lez}, P.~G., {Rieke}, G.~H.,
  {Alexander}, D.~M., {Rigby}, J.~R., {Papovich}, C., {Donley}, J.~L., \&
  {Rigopoulou}, D. 2008, \apj, 677, 127

\bibitem[Arnaud(1996)]{arnaud96} Arnaud, K.~A.\ 1996, 
Astronomical Data Analysis Software and Systems V, 101, 17 

\bibitem[{{Baldry} {et~al.}(2004){Baldry}, {Glazebrook}, {Brinkmann},
  {Ivezi{\'c}}, {Lupton}, {Nichol}, \& {Szalay}}]{baldry04}
{Baldry}, I.~K., {Glazebrook}, K., {Brinkmann}, J., {Ivezi{\'c}}, {\v Z}.,
  {Lupton}, R.~H., {Nichol}, R.~C., \& {Szalay}, A.~S. 2004, \apj, 600, 681

\bibitem[{{Ballantyne} {et~al.}(2006){Ballantyne}, {Everett}, \&
  {Murray}}]{ballantyne06}
{Ballantyne}, D.~R., {Everett}, J.~E., \& {Murray}, N. 2006, \apj, 639, 740

\bibitem[{{Barger} {et~al.}(2003){Barger}, {Cowie}, {Capak}, {Alexander},
  {Bauer}, {Fernandez}, {Brandt}, {Garmire}, \& {Hornschemeier}}]{barger03}
{Barger}, A.~J., {Cowie}, L.~L., {Capak}, P., {Alexander}, D.~M., {Bauer},
  F.~E., {Fernandez}, E., {Brandt}, W.~N., {Garmire}, G.~P., \&
  {Hornschemeier}, A.~E. 2003, \aj, 126, 632

\bibitem[{{Barger} {et~al.}(2005){Barger}, {Cowie}, {Mushotzky}, {Yang},
  {Wang}, {Steffen}, \& {Capak}}]{barger05}
{Barger}, A.~J., {Cowie}, L.~L., {Mushotzky}, R.~F., {Yang}, Y., {Wang}, W.-H.,
  {Steffen}, A.~T., \& {Capak}, P. 2005, \aj, 129, 578

\bibitem[{{Barvainis}(1987)}]{barvainis87}
{Barvainis}, R. 1987, \apj, 320, 537

\bibitem[{{Bauer} {et~al.}(2004){Bauer}, {Alexander}, {Brandt}, {Schneider},
  {Treister}, {Hornschemeier}, \& {Garmire}}]{bauer04}
{Bauer}, F.~E., {Alexander}, D.~M., {Brandt}, W.~N., {Schneider}, D.~P.,
  {Treister}, E., {Hornschemeier}, A.~E., \& {Garmire}, G.~P. 2004, \aj, 128,
  2048

\bibitem[{{Bell} {et~al.}(2004){Bell}, {Wolf}, {Meisenheimer}, {Rix}, {Borch},
  {Dye}, {Kleinheinrich}, {Wisotzki}, \& {McIntosh}}]{bell04}
{Bell}, E.~F., {Wolf}, C., {Meisenheimer}, K., {Rix}, H.-W., {Borch}, A.,
  {Dye}, S., {Kleinheinrich}, M., {Wisotzki}, L., \& {McIntosh}, D.~H. 2004,
  \apj, 608, 752

\bibitem[{{Brandt} {et~al.}(2001){Brandt}, {Alexander}, {Hornschemeier},
  {Garmire}, {Schneider}, {Barger}, {Bauer}, {Broos}, {Cowie}, {Townsley},
  {Burrows}, {Chartas}, {Feigelson}, {Griffiths}, {Nousek}, \&
  {Sargent}}]{brandt01}
{Brandt}, W.~N., {Alexander}, D.~M., {Hornschemeier}, A.~E., {Garmire}, G.~P.,
  {Schneider}, D.~P., {Barger}, A.~J., {Bauer}, F.~E., {Broos}, P.~S., {Cowie},
  L.~L., {Townsley}, L.~K., {Burrows}, D.~N., {Chartas}, G., {Feigelson},
  E.~D., {Griffiths}, R.~E., {Nousek}, J.~A., \& {Sargent}, W.~L.~W. 2001, \aj,
  122, 2810

\bibitem[{{Cardamone} {et~al.}(2007){Cardamone}, {Moran}, \&
  {Kay}}]{cardamone07}
{Cardamone}, C.~N., {Moran}, E.~C., \& {Kay}, L.~E. 2007, \aj, 134, 1263

\bibitem[{{Cardamone} {et~al.}(2008){Cardamone}, {Urry}, {Damen}, {van Dokkum},
  {Treister}, {Labb{\'e}}, {Virani}, {Lira}, \& {Gawiser}}]{cardamone08}
{Cardamone}, C.~N., {Urry}, C.~M., {Damen}, M., {van Dokkum}, P., {Treister},
  E., {Labb{\'e}}, I., {Virani}, S.~N., {Lira}, P., \& {Gawiser}, E. 2008,
  \apj, 680, 130

\bibitem[{{Cirasuolo} {et~al.}(2007){Cirasuolo}, {McLure}, {Dunlop}, {Almaini},
  {Foucaud}, {Smail}, {Sekiguchi}, {Simpson}, {Eales}, {Dye}, {Watson}, {Page},
  \& {Hirst}}]{cirasuolo07}
{Cirasuolo}, M., {McLure}, R.~J., {Dunlop}, J.~S., {Almaini}, O., {Foucaud},
  S., {Smail}, I., {Sekiguchi}, K., {Simpson}, C., {Eales}, S., {Dye}, S.,
  {Watson}, M.~G., {Page}, M.~J., \& {Hirst}, P. 2007, \mnras, 380, 585

\bibitem[{{Comastri} {et~al.}(1995){Comastri}, {Setti}, {Zamorani}, \&
  {Hasinger}}]{comastri95}
{Comastri}, A., {Setti}, G., {Zamorani}, G., \& {Hasinger}, G. 1995, \aap, 296,
  1

\bibitem[{{Croom} {et~al.}(2001){Croom}, {Warren}, \& {Glazebrook}}]{croom01}
{Croom}, S.~M., {Warren}, S.~J., \& {Glazebrook}, K. 2001, \mnras, 328, 150

\bibitem[{{Davis} {et~al.}(2007)}]{davis07}
{Davis}, M. {et~al.} 2007, \apjl, 660, L1

\bibitem[{{De Luca} \& {Molendi}(2004)}]{deluca04}
{De Luca}, A. \& {Molendi}, S. 2004, \aap, 419, 837

\bibitem[{{Della Ceca} {et~al.}(2008){Della Ceca}, {Caccianiga}, {Severgnini},
  {Maccacaro}, {Brunner}, {Carrera}, {Cocchia}, {Mateos}, {Page}, \&
  {Tedds}}]{dellaceca08}
{Della Ceca}, R., {Caccianiga}, A., {Severgnini}, P., {Maccacaro}, T.,
  {Brunner}, H., {Carrera}, F.~J., {Cocchia}, F., {Mateos}, S., {Page}, M.~J.,
  \& {Tedds}, J.~A. 2008, ArXiv e-prints, 805

\bibitem[{{Fazio} {et~al.}(2004)}]{fazio04a}
{Fazio}, G.~G. {et~al.} 2004, \apjs, 154, 10

\bibitem[{{Gawiser} {et~al.}(2006{\natexlab{a}})}]{gawiser06a}
{Gawiser}, E. {et~al.} 2006{\natexlab{a}}, \apjs, 162, 1

\bibitem[{{Gawiser} {et~al.}(2006{\natexlab{b}})}]{gawiser06b}
---. 2006{\natexlab{b}}, \apjl, 642, L13

\bibitem[Georgakakis et al.(2008)]{georgakakis08} Georgakakis, A., et 
al.\ 2008, \mnras, 385, 2049

\bibitem[{{Giacconi} {et~al.}(2001){Giacconi}, {Rosati}, {Tozzi}, {Nonino},
  {Hasinger}, {Norman}, {Bergeron}, {Borgani}, {Gilli}, {Gilmozzi}, \&
  {Zheng}}]{giacconi01}
{Giacconi}, R., {Rosati}, P., {Tozzi}, P., {Nonino}, M., {Hasinger}, G.,
  {Norman}, C., {Bergeron}, J., {Borgani}, S., {Gilli}, R., {Gilmozzi}, R., \&
  {Zheng}, W. 2001, \apj, 551, 624

\bibitem[{{Gilli} {et~al.}(2007){Gilli}, {Comastri}, \& {Hasinger}}]{gilli07}
{Gilli}, R., {Comastri}, A., \& {Hasinger}, G. 2007, \aap, 463, 79

\bibitem[{{Gilli} {et~al.}(2005){Gilli}, {Daddi}, {Zamorani}, {Tozzi},
  {Borgani}, {Bergeron}, {Giacconi}, {Hasinger}, {Mainieri}, {Norman},
  {Rosati}, {Szokoly}, \& {Zheng}}]{gilli05}
{Gilli}, R., {Daddi}, E., {Zamorani}, G., {Tozzi}, P., {Borgani}, S.,
  {Bergeron}, J., {Giacconi}, R., {Hasinger}, G., {Mainieri}, V., {Norman}, C.,
  {Rosati}, P., {Szokoly}, G., \& {Zheng}, W. 2005, \aap, 430, 811

\bibitem[{{Gilli} {et~al.}(2001){Gilli}, {Salvati}, \& {Hasinger}}]{gilli01}
{Gilli}, R., {Salvati}, M., \& {Hasinger}, G. 2001, \aap, 366, 407

\bibitem[{{Gilli} {et~al.}(2003)}]{gilli03}
{Gilli}, R. {et~al.} 2003, \apj, 592, 721

\bibitem[{{Gruber}(1992)}]{gruber92}
{Gruber}, D.~E. 1992, in The X-ray Background, 44

\bibitem[{{Hasinger} {et~al.}(2005){Hasinger}, {Miyaji}, \&
  {Schmidt}}]{hasinger05}
{Hasinger}, G., {Miyaji}, T., \& {Schmidt}, M. 2005, \aap, 441, 417

\bibitem[{{Hickox} \& {Markevitch}(2006)}]{hickox06}
{Hickox}, R.~C. \& {Markevitch}, M. 2006, \apj, 645, 95

\bibitem[{{Hildebrandt} {et~al.}(2006){Hildebrandt}, {Erben}, {Dietrich},
  {Cordes}, {Haberzettl}, {Hetterscheidt}, {Schirmer}, {Schmithuesen},
  {Schneider}, {Simon}, \& {Trachternach}}]{hildebrandt06}
{Hildebrandt}, H., {Erben}, T., {Dietrich}, J.~P., {Cordes}, O., {Haberzettl},
  L., {Hetterscheidt}, M., {Schirmer}, M., {Schmithuesen}, O., {Schneider}, P.,
  {Simon}, P., \& {Trachternach}, C. 2006, \aap, 452, 1121

\bibitem[{{H{\"o}nig} \& {Beckert}(2007)}]{hoenig07}
{H{\"o}nig}, S.~F. \& {Beckert}, T. 2007, \mnras, 380, 1172

\bibitem[{{H{\"o}nig} {et~al.}(2006){H{\"o}nig}, {Beckert}, {Ohnaka}, \&
  {Weigelt}}]{hoenig06}
{H{\"o}nig}, S.~F., {Beckert}, T., {Ohnaka}, K., \& {Weigelt}, G. 2006, \aap,
  452, 459

\bibitem[{{Horst} {et~al.}(2008){Horst}, {Gandhi}, {Smette}, \&
  {Duschl}}]{horst08}
{Horst}, H., {Gandhi}, P., {Smette}, A., \& {Duschl}, W.~J. 2008, \aap, 479,
  389

\bibitem[{{Horst} {et~al.}(2006){Horst}, {Smette}, {Gandhi}, \&
  {Duschl}}]{horst06}
{Horst}, H., {Smette}, A., {Gandhi}, P., \& {Duschl}, W.~J. 2006, \aap, 457,
  L17

\bibitem[{{Kalberla} {et~al.}(2005){Kalberla}, {Burton}, {Hartmann}, {Arnal},
  {Bajaja}, {Morras}, \& {P{\"o}ppel}}]{kalberla05}
{Kalberla}, P.~M.~W., {Burton}, W.~B., {Hartmann}, D., {Arnal}, E.~M.,
  {Bajaja}, E., {Morras}, R., \& {P{\"o}ppel}, W.~G.~L. 2005, \aap, 440, 775

\bibitem[{{Krolik} \& {Begelman}(1988)}]{krolik88}
{Krolik}, J.~H. \& {Begelman}, M.~C. 1988, \apj, 329, 702

\bibitem[{{La Franca} {et~al.}(2005)}]{lafranca05}
{La Franca}, F. {et~al.} 2005, \apj, 635, 864

\bibitem[La Franca \& Cristiani(1997)]{lafranca97} La Franca, F., \& Cristiani, S.\ 1997, \aj, 113, 1517 

\bibitem[{{Lawrence}(1991)}]{lawrence91}
{Lawrence}, A. 1991, \mnras, 252, 586

\bibitem[{{Le F{\` e}vre} {et~al.}(2004)}]{lefevre04}
{Le F{\` e}vre}, O. {et~al.} 2004, \aap, 428, 1043

\bibitem[{{LeFevre} {et~al.}(2003)}]{lefevre03}
{LeFevre}, O. {et~al.} 2003, in Presented at the Society of Photo-Optical
  Instrumentation Engineers (SPIE) Conference, Vol. 4841, Instrument Design and
  Performance for Optical/Infrared Ground-based Telescopes. Edited by Iye,
  Masanori; Moorwood, Alan F. M. Proceedings of the SPIE, Volume 4841, pp.
  1670-1681 (2003)., ed. M.~{Iye} \& A.~F.~M. {Moorwood}, 1670--1681

\bibitem[{{Lehmer} {et~al.}(2005)}]{lehmer05}
{Lehmer}, B.~D. {et~al.} 2005, \apjs, 161, 21

\bibitem[{{Lira} {et~al.}(2002){Lira}, {Ward}, {Zezas}, {Alonso-Herrero}, \&
  {Ueno}}]{lira02}
{Lira}, P., {Ward}, M., {Zezas}, A., {Alonso-Herrero}, A., \& {Ueno}, S. 2002,
  \mnras, 330, 259

\bibitem[{{Lutz} {et~al.}(2004){Lutz}, {Maiolino}, {Spoon}, \&
  {Moorwood}}]{lutz04}
{Lutz}, D., {Maiolino}, R., {Spoon}, H.~W.~W., \& {Moorwood}, A.~F.~M. 2004,
  \aap, 418, 465

\bibitem[{{Makovoz} \& {Marleau}(2005)}]{makovoz05}
{Makovoz}, D. \& {Marleau}, F.~R. 2005, \pasp, 117, 1113

\bibitem[{{Mignano} {et~al.}(2007){Mignano}, {Miralles}, {da Costa}, {Olsen},
  {Prandoni}, {Arnouts}, {Benoist}, {Madejsky}, {Slijkhuis}, \&
  {Zaggia}}]{mignano07}
{Mignano}, A., {Miralles}, J.-M., {da Costa}, L., {Olsen}, L.~F., {Prandoni},
  I., {Arnouts}, S., {Benoist}, C., {Madejsky}, R., {Slijkhuis}, R., \&
  {Zaggia}, S. 2007, \aap, 462, 553

\bibitem[{{Moran} {et~al.}(2002){Moran}, {Filippenko}, \& {Chornock}}]{moran02}
{Moran}, E.~C., {Filippenko}, A.~V., \& {Chornock}, R. 2002, \apjl, 579, L71

\bibitem[{{Morrison} \& {McCammon}(1983)}]{morrison83}
{Morrison}, R. \& {McCammon}, D. 1983, \apj, 270, 119

\bibitem[{{Nandra} {et~al.}(2007){Nandra}, {Georgakakis}, {Willmer}, {Cooper},
  {Croton}, {Davis}, {Faber}, {Koo}, {Laird}, \& {Newman}}]{nandra07}
{Nandra}, K., {Georgakakis}, A., {Willmer}, C.~N.~A., {Cooper}, M.~C.,
  {Croton}, D.~J., {Davis}, M., {Faber}, S.~M., {Koo}, D.~C., {Laird}, E.~S.,
  \& {Newman}, J.~A. 2007, \apjl, 660, L11

\bibitem[{{Nandra} {et~al.}(1997){Nandra}, {George}, {Mushotzky}, {Turner}, \&
  {Yaqoob}}]{nandra97}
{Nandra}, K., {George}, I.~M., {Mushotzky}, R.~F., {Turner}, T.~J., \&
  {Yaqoob}, T. 1997, \apj, 476, 70

\bibitem[{{Nandra} \& {Pounds}(1994)}]{nandra94}
{Nandra}, K. \& {Pounds}, K.~A. 1994, \mnras, 268, 405

\bibitem[Nenkova et al.(2008)]{nenkova08} Nenkova, M., Sirocky, 
M.~M., Nikutta, R., Ivezi{\'c}, {\v Z}., 
\& Elitzur, M.\ 2008, \apj, 685, 160 

\bibitem[{{Nenkova} {et~al.}(2002){Nenkova}, {Ivezi{\' c}}, \&
  {Elitzur}}]{nenkova02}
{Nenkova}, M., {Ivezi{\' c}}, {\v Z}., \& {Elitzur}, M. 2002, \apjl, 570, L9

\bibitem[{{Oke} \& {Gunn}(1983)}]{oke83}
{Oke}, J.~B. \& {Gunn}, J.~E. 1983, \apj, 266, 713

\bibitem[{{Pier} \& {Krolik}(1992)}]{pier92}
{Pier}, E.~A. \& {Krolik}, J.~H. 1992, \apj, 401, 99

\bibitem[{{Pier} \& {Krolik}(1993)}]{pier93}
---. 1993, \apj, 418, 673

\bibitem[{{Polletta} {et~al.}(2007)}]{polletta07}
{Polletta}, M. {et~al.} 2007, \apj, 663, 81

\bibitem[{{Ramos Almeida} {et~al.}(2007){Ramos Almeida}, {P{\'e}rez
  Garc{\'{\i}}a}, {Acosta-Pulido}, \& {Rodr{\'{\i}}guez
  Espinosa}}]{ramos-almeida07}
{Ramos Almeida}, C., {P{\'e}rez Garc{\'{\i}}a}, A.~M., {Acosta-Pulido}, J.~A.,
  \& {Rodr{\'{\i}}guez Espinosa}, J.~M. 2007, \aj, 134, 2006

\bibitem[{{Rieke} {et~al.}(2004)}]{rieke04}
{Rieke}, G.~H. {et~al.} 2004, \apjs, 154, 25

\bibitem[{{Risaliti} {et~al.}(1999){Risaliti}, {Maiolino}, \&
  {Salvati}}]{risaliti99}
{Risaliti}, G., {Maiolino}, R., \& {Salvati}, M. 1999, \apj, 522, 157

\bibitem[{{Rosati} {et~al.}(2002)}]{rosati02}
{Rosati}, P. {et~al.} 2002, \apj, 566, 667

\bibitem[{{Schawinski} {et~al.}(2008)}]{schawinski08}
{Schawinski}, K., {Virani}, S., {Simmons}, B., {Urry}, C.M., {Treister}, E. \&
{Kaviraj}, S. 2008, \apj, submitted

\bibitem[{{Schawinski} {et~al.}(2006)}]{schawinski06}
{Schawinski}, K. {et~al.} 2006, \nat, 442, 888

\bibitem[{{Schneider} {et~al.}(2002)}]{schneider02}
{Schneider}, D.~P. {et~al.} 2002, \aj, 123, 567

\bibitem[{{Scoville} {et~al.}(2007)}]{scoville07}
{Scoville}, N. {et~al.} 2007, \apjs, 172, 1

\bibitem[{{Setti} \& {Woltjer}(1989)}]{setti89}
{Setti}, G. \& {Woltjer}, L. 1989, \aap, 224, L21

\bibitem[{{Silverman} {et~al.}(2008)}]{silverman08}
{Silverman}, J.~D. {et~al.} 2008, \apj, 675, 1025

\bibitem[{{Simpson}(2005)}]{simpson05}
{Simpson}, C. 2005, \mnras, 360, 565

\bibitem[{{Spergel} {et~al.}(2007)}]{spergel07}
{Spergel}, D.~N. {et~al.} 2007, \apjs, 170, 377

\bibitem[{{Springel} {et~al.}(2005){Springel}, {Di Matteo}, \&
  {Hernquist}}]{springel05}
{Springel}, V., {Di Matteo}, T., \& {Hernquist}, L. 2005, \mnras, 361, 776

\bibitem[{{Steffen} {et~al.}(2003){Steffen}, {Barger}, {Cowie}, {Mushotzky}, \&
  {Yang}}]{steffen03}
{Steffen}, A.~T., {Barger}, A.~J., {Cowie}, L.~L., {Mushotzky}, R.~F., \&
  {Yang}, Y. 2003, \apjl, 596, L23

\bibitem[{{Szokoly} {et~al.}(2004)}]{szokoly04}
{Szokoly}, G.~P. {et~al.} 2004, \apjs, 155, 271

\bibitem[{{Tozzi} {et~al.}(2006)}]{tozzi06}
{Tozzi}, P. {et~al.} 2006, \aap, 451, 457

\bibitem[{{Treister} {et~al.}(2005){Treister}, {Castander}, {Maccarone},
  {Gawiser}, {Coppi}, {Urry}, {Maza}, {Herrera}, {Gonzalez}, {Montoya}, \&
  {Pineda}}]{treister05a}
{Treister}, E., {Castander}, F.~J., {Maccarone}, T.~J., {Gawiser}, E., {Coppi},
  P.~S., {Urry}, C.~M., {Maza}, J., {Herrera}, D., {Gonzalez}, V., {Montoya},
  C., \& {Pineda}, P. 2005, \apj, 621, 104

\bibitem[{{Treister} {et~al.}(2007){Treister}, {Gawiser}, {van Dokkum}, {Lira},
  {Urry}, \& {The Musyc Collaboration}}]{treister07}
{Treister}, E., {Gawiser}, E., {van Dokkum}, P., {Lira}, P., {Urry}, M., \&
  {The Musyc Collaboration}. 2007, The Messenger, 129, 45

\bibitem[{{Treister} {et~al.}(2008){Treister}, {Krolik}, \&
  {Dullemond}}]{treister08}
{Treister}, E., {Krolik}, J.~H., \& {Dullemond}, C. 2008, \apj, 679, 140

\bibitem[{{Treister} \& {Urry}(2005)}]{treister05b}
{Treister}, E. \& {Urry}, C.~M. 2005, \apj, 630, 115

\bibitem[{{Treister} \& {Urry}(2006)}]{treister06b}
---. 2006, \apjl, 652, L79

\bibitem[{{Treister} {et~al.}(2006){Treister}, {Urry}, {Van Duyne},
  {Dickinson}, {Chary}, {Alexander}, {Bauer}, {Natarajan}, {Lira}, \&
  {Grogin}}]{treister06a}
{Treister}, E., {Urry}, C.~M., {Van Duyne}, J., {Dickinson}, M., {Chary},
  R.-R., {Alexander}, D.~M., {Bauer}, F., {Natarajan}, P., {Lira}, P., \&
  {Grogin}, N.~A. 2006, \apj, 640, 603

\bibitem[{{Treister} {et~al.}(2004)}]{treister04}
{Treister}, E. {et~al.} 2004, \apj, 616, 123

\bibitem[{{Ueda} {et~al.}(2003){Ueda}, {Akiyama}, {Ohta}, \& {Miyaji}}]{ueda03}
{Ueda}, Y., {Akiyama}, M., {Ohta}, K., \& {Miyaji}, T. 2003, \apj, 598, 886

\bibitem[{{Vanzella} {et~al.}(2005)}]{vanzella05}
{Vanzella}, E. {et~al.} 2005, \aap, 434, 53

\bibitem[{{Vanzella} {et~al.}(2006)}]{vanzella06}
---. 2006, \aap, 454, 423

\bibitem[{{Vanzella} {et~al.}(2008)}]{vanzella08}
---. 2008, \aap, 478, 83

\bibitem[{{Virani} {et~al.}(2006){Virani}, {Treister}, {Urry}, \&
  {Gawiser}}]{virani06}
{Virani}, S.~N., {Treister}, E., {Urry}, C.~M., \& {Gawiser}, E. 2006, \aj,
  131, 2373

\bibitem[{{Weiner} {et~al.}(2005)}]{weiner05}
{Weiner}, B.~J. {et~al.} 2005, \apj, 620, 595

\bibitem[{{Wolf} {et~al.}(2004)}]{wolf04}
{Wolf}, C. {et~al.} 2004, \aap, 421, 913

\bibitem[{{Zheng} {et~al.}(2004)}]{zheng04}
{Zheng}, W. {et~al.} 2004, \apjs, 155, 73

\end{thebibliography}

\newpage

\begin{deluxetable}{lccccccc}
\tablecolumns{8}
\tabletypesize{\scriptsize}
\tablecaption{Log of Spectroscopic Observations}
\tablehead{
\colhead{Run} & \colhead{Instrument} & \colhead{Masks} & \colhead{Slit Width} & \colhead{Avg. Seeing} & \colhead{Total Sources} & \colhead{X-ray Sources} & \colhead{Efficiency\tablenotemark{a}} }
\startdata
26-27/10/2003                 & IMACS & 4 & 1.0$''$ & $\sim$1$''$   & 291 & 74  & 57\%\\
4-7/2/2005                    & IMACS & 2 & 1.2$''$ & $\sim$0.8$''$ & 180 & 64  & 57\%\\
2-3/11/2005                   & IMACS & 2 & 1.2$''$ & $\sim$0.6$''$ & 194 & 29  & 66\%\\
25-27/10/2006                 & IMACS & 3 & 1.0$''$ & 0.5-1.5$''$   & 280 & 143 & 32\%\\
21-22/11/2006                 & IMACS & 1 & 1.2$''$ & $\sim$0.8$''$ & 109 & 17  & 20\%\\
18-20/2/2007\tablenotemark{b} & IMACS & 1 & 1.2$''$ & $\sim$1$''$   & 109 & 17  & 20\%\\
Period 78                     & VIMOS & 4 & 1.0$''$ & 1$''$         & 283 & 96  & 54\%\\
\enddata
\tablenotetext{a}{Defined as the fraction of X-ray sources identified.} 
\tablenotetext{b}{The mask from the Nov. 2006 run was re-observed in order to obtain $\sim$5 hours of integration time.}
\label{log_imacs}
\end{deluxetable}

\clearpage
\begin{landscape}
\begin{deluxetable}{lcccccccccccccccc}
\tablecolumns{17}
\tabletypesize{\scriptsize}
\tablewidth{0pc}
\tablecaption{\label{cat}X-ray, optical and mid-IR properties of the X-ray sources in the ECDF-S}
\tablehead{
\colhead{ID\tablenotemark{a}} & \multicolumn{2}{c}{X-ray Flux (erg~cm$^2$s$^{-1}$)} & \multicolumn{3}{c}{Hardness Ratio} & \colhead{Redshift} & \colhead{Instrument} & \colhead{Class.\tablenotemark{d}}
& \colhead{log (X-ray Lum.)} & \multicolumn{3}{c}{24~$\mu$m (mJy)} & \multicolumn{3}{c}{$\Gamma$}  & \colhead{$N_H$}\\
\colhead{} & \colhead{Soft\tablenotemark{b}} & \colhead{Hard\tablenotemark{c}} & \colhead{Value} & \colhead{Upper} & \colhead{Lower} & \colhead{} & \colhead{} & \colhead{}
& \colhead{(erg~s$^{-1}$)} & \colhead{Flux} & \colhead{Error} & \colhead{Source\tablenotemark{e}} & \colhead{Value} & \colhead{Upper} & \colhead{Lower} & \colhead{(cm$^{-2}$)}
}
\startdata
  2 & 2.040e-15 & 4.690e-15 &  1.000 & 0.000 & 0.000 & 9.990\tablenotemark{f} & IMACS & UNK  & ------ &   51.27 &  1.42 & F & ----- & ----- & ----- & ------\\
  3 & 2.670e-15 & 6.620e-15 & -1.000 & 0.000 & 0.000 & 9.990 & IMACS & UNK  & ------ &  223.50 &  1.33 & F & 1.900 & ----- & ----- & 20.843\\
  4 & 3.650e-14 & 6.590e-14 & -0.354 & 0.030 & 0.031 & 2.011 & IMACS & OAGN & 45.273 &  229.90 &  1.60 & F & 1.836 & 0.123 & 0.117 & 20.941\\
  6 & 6.760e-15 & 8.590e-15 & -0.498 & 0.069 & 0.075 & 3.031 & VIMOS & UAGN & 44.821 &  482.20 &  1.22 & F & 2.045 & 0.402 & 0.347 & 21.087\\
  7 & 1.260e-15 & 4.300e-15 & -0.054 & 0.126 & 0.146 & 0.289 & IMACS & ELG  & 42.044 &   92.09 &  1.26 & F & ----- & ----- & ----- & ------\\
\enddata
\tablenotetext{a}{X-ray ID number from the catalog of \citet{virani06}}
\tablenotetext{b}{0.5--2~keV}
\tablenotetext{c}{2--8~keV}
\tablenotetext{d}{UNK: Unkown, OAGN: Obscured AGN, UAGN: Unobuscred AGN, ELG: Emission line galaxy, ALG: Absorption line galaxy, STAR: star.}
\tablenotetext{e}{Survey in which the 24~$\mu$m source is identified. F: FIDEL, G: GOODS.}
\tablenotetext{f}{A $z$=9.990 corresponds to a targeted source for which the redshift could not be measured.}
\tablecomments{This table is published in its entirety in the electronic edition of the Astrophysical Journal. A portion is shown here for 
guidance regarding its form and content.}
\end{deluxetable}
\clearpage
\end{landscape}

\end{document}